\documentclass[compsoc,onecolumn,letter]{IEEEtran}
\usepackage{mathptmx} 

\newcommand{\ignore}[1]{}
\usepackage{array}
\newcolumntype{?}{!{\vrule width 2pt}}
\usepackage{mathtools}
\usepackage{array}
\usepackage{fancyhdr}
\usepackage[normalem]{ulem}
\usepackage[hyphens]{url}
\usepackage[sort,nocompress]{cite}
\usepackage[final]{microtype}
\usepackage{flushend}
\usepackage{etoolbox}

\usepackage{hyperref}

\usepackage{comment}
\usepackage[]{subfig}

\usepackage{float}

\usepackage{multirow}
\usepackage{hhline}

\usepackage{makecell}
\usepackage{xcolor}

\usepackage{tabularx}
\usepackage{graphicx}
\usepackage[capitalise]{cleveref}
\usepackage[font={small}]{caption}
\usepackage{url}

\usepackage{xspace}

\newcommand{\OURLCORE}{TensorDash} 
\newcommand{\OURSCORE}{TensorDash}
\newcommand{\OURL}{\textit{\OURLCORE}\xspace} 
\newcommand{\OURS}{\textit{\OURSCORE}\xspace} 

\DeclareRobustCommand{\Diffy}{$\mathit{Diffy}$\xspace}

\usepackage{ragged2e}
\usepackage{changepage}

\usepackage{color,soul}
\soulregister\cref7
\soulregister\Diffy7
\soulregister\textsubscript7
\soulregister\cite7
\definecolor{lightgreen}{RGB}{195, 233, 211}
\definecolor{lightred}{RGB}{233, 0, 0}
\definecolor{white}{RGB}{255, 255, 255}
\sethlcolor{lightgreen}

\newcommand{\iscasubmissionnumber}{560}

\fancypagestyle{firstpage}{
  \fancyhf{}

  \fancyhead[C]{\normalsize{ISCA 2020 Submission
      \textbf{\#\iscasubmissionnumber} \\ Confidential Draft: DO NOT DISTRIBUTE}} 
  \fancyfoot[C]{\thepage}
}

\title{\OURLCORE: Exploiting Sparsity to Accelerate Deep Neural Network Training and Inference} 
\author{\IEEEauthorblockN{Mostafa Mahmoud\textsuperscript{1}, Isak Edo\textsuperscript{1}, Ali Hadi Zadeh\textsuperscript{1},  Omar Mohamed Awad\textsuperscript{1},\\
Gennady Pekhimenko\textsuperscript{1,3}, Jorge Albericio\textsuperscript{2} and Andreas Moshovos\textsuperscript{1,3}}
\IEEEauthorblockA{ 
\\
1. University of Toronto,  2. Cerebras Systems, 3. Vector Institute \\
\{mostafa.mahmoud, isak.edo, a.hadizadeh, omar.awad\}@mail.utoronto.ca,\\
pekhimenko@cs.toronto.edu, jorge@cerebras.net, moshovos@ece.utoronto.ca}
}

\begin{document}

\maketitle

\begin{abstract}
\justify
\begin{adjustwidth}{1cm}{1cm}\noindent\textbf{Abstract: }\OURL is a hardware level technique for enabling data-parallel MAC units to take advantage of sparsity in their input operand streams. When used to compose a hardware  accelerator for deep learning, \OURL can speedup the \textit{training} process while also increasing energy efficiency. \OURL combines a low-cost, sparse input operand interconnect comprising an 8-input multiplexer per multiplier input, with an area efficient hardware scheduler. While the interconnect allows a very limited set of movements per operand, the scheduler can effectively extract sparsity when it is present in the activations, weights or gradients of neural networks. Over a wide set of models covering various applications, \OURS accelerates the training process by $1.95{\times}$ while being $1.89\times$ more energy efficient, $1.6\times$ more energy efficient when taking on-chip and off-chip memory accesses into account. While \OURL works with any datatype, we demonstrate it with both single-precision floating-point units and bfloat16. 
\end{adjustwidth}

\end{abstract}

\section{Introduction}
\label{sec:intro}
Neural networks are being used in an ever increasing number of application domains delivering state-of-the-art results. Given their high computation and memory demands and their increasing importance, considerable attention has also been given into techniques for optimizing implementations at all system levels all the way down to specialized hardware. Whereas a decade ago the then state-of-the-art neural networks could be trained on a commodity server within a few hours, today training the best neural network models has become an exascale class problem~\cite{ScaleDeep}. State-of-the-art neural networks now require many graphics processors or specialized accelerators such as the TPU~\cite{TPU} so that they can be trained within practical time limits. Tuning neural networks for best performance during inference further exacerbates the cost of training. Beyond the cost of acquiring or getting access to such expensive computing resources, worse are the operating costs and the environmental impact of training. Strubell et al., report that the CO\textsubscript{2} emissions of training even a mid-class neural network stand at about 36 metric tons which is more than double the estimated 16.5 metric tons needed on average per person and per year in the US~\cite{ML_co2}. Training neural networks at the ``edge'' is needed in certain applications such as for example to refine an existing model with user-specific information and input. While the trade offs for edge devices are different than those in data centers or desktop applications, the need remains the same: reduce execution time and improve energy efficiency albeit under different constraints.

It comes then as no surprise that efforts for reducing the execution time and the energy cost of training have been considerable. First and foremost, by exploiting model, data, and pipeline parallelism distributed training partitions the training workload across several computing nodes to reduce overall latency~\cite{dist_training1,dist_training2,dist_training3}. Intra- and inter-node data blocking, reuse, and communication and computation overlapping orchestrate the use of the computing, memory hierarchy, and communication resources to improve performance and energy efficiency~\cite{eyeriss, TernGrad,UCNN}. Lossless and lossy compression reduces the footprint of the vast amounts of data processed during training~\cite{GIST}. While originally training used double precision floating-point data and arithmetic, more compact datatypes reduce overall data volumes and computation costs. These include: single precision floating-point, bfloat16~\cite{bfloat16Google,bfloat16,bfloat16_2}, dynamic floating-point~\cite{DBLP:conf/iclr/0002MMKAB0VKGHD18}, and flexpoint~\cite{Koster:2017:FAN:3294771.3294937}. Mixed-datatype methods further reduce costs by performing many computations using fixed-point and few using some form of floating-point~\cite{DBLP:conf/iclr/0002MMKAB0VKGHD18, mixedP, nvidia_mixedP, Drumond:2018:TDH:3326943.3326985}. Other methods use low precision arithmetic~\cite{HALP}.

Even with these techniques training remains an exascale class problem and further improvements are needed. Accordingly, in this work we are proposing a technique for further improving execution time and energy efficiency for training. Specifically, we propose \OURL exploits ineffectual operations that occur \textit{naturally}  for many models during training. The bulk of the energy during training is due to the transfers and computations needed to perform multiply-accumulate operations (MACs). We find that often one of the operands in these MACs is zero. These operations can be safely eliminated as they do not affect the values produced during training and thus convergence and final accuracy. We find that for \textit{many} networks a large number of zeros naturally occur in the activation values during the forward and backward passes, and in the gradients during the backward pass (see Section~\ref{sec:training_overview} for a primer on training). When sparsity exists it represents an opportunity for improving performance and energy efficiency. Accordingly, we seek to develop a method that will do so when sparsity exists and that will not hurt performance and energy efficiency otherwise. 

The sparsity pattern during training is \textit{dynamic}. It changes with the input and varies across epochs and batches. Accordingly, \OURL uses a run-time approach where the elimination of ineffectual MACs is performed using a combination of an inexpensive hardware scheduler and a co-designed sparse, low-cost data interconnect that are placed just in front of the MAC units. \OURL not only eliminated ineffectual MACs but it also advances in their place other effectual MACs that would otherwise have executed later in time. This improves energy efficiency and performance. \OURL works with out-of-the-box neural networks and requires no modification nor any special annotations from the model developer. It simply extracts and exploits naturally occurring sparsity regardless of how it is distributed. 

More importantly, \OURL extracts additional benefits from another class of existing training acceleration techniques: These are techniques that perform network pruning and quantization during training. Pruning's goal is to convert weight values to zero. As training proceeds with pruning, we observe that pruning results in increased sparsity not only in the weights but also in the activations and the gradients.  Quantization's goal is to reduce the datawidth that will be used during inference. During training quantization effectively clips what would otherwise be values of low magnitude into zeros.  Dynamic sparse reparameterization \cite{dynamic_sparse_reparam}, eager pruning \cite{eager_pruning} and DropBack \cite{DropBack}, and PACT \cite{PACT} and LQ-Nets \cite{LQ-nets} are examples of recent training-time pruning, and quantization techniques respectively. We study the interaction of \OURL and some of these methods. \OURL would also benefit selective backpropagation methods which backpropagate loss only for some of the neurons~\cite{MeProp}. Unless specialized hardware is developed, selective backpropagation manifests as sparsity as it effectively converts a large number of gradients into zeros.

Our contribution is that we propose \OURL with the following functionality and benefits:

\begin{itemize}
\item \OURL exploits naturally occurring sparsity during training which appears predominantly in the activations and the gradients.
\item \OURL exploits sparsity dynamically and completely in hardware. It utilizes a low-overhead hardware scheduler to advance MAC operations in time (earlier cycle) and space (MAC unit) so that overall computation finishes earlier. The scheduler makes no assumptions about how sparsity is distributed so that it can handle the dynamic sparsity patterns that arise during training.
\item \OURL does not affect numerical fidelity. It only eliminates MAC operations where at least one of the inputs is zero.
\item \OURL is compatible with data-parallel processing elements that perform multiple MAC operations all accumulating into a single value and is compatible with any dataflow for such processing elements.
\item Benefits with \OURL are amplified with training algorithms that incorporate quantization, pruning and selective backpropagation.
\item \OURL would also benefit inference.
\item The core processing element \OURL uses can be configured to extract sparsity on one or both operands. For training we configure it to do so only on one side as this proves sufficient.
\item For models where sparsity is insufficient \OURL could automatically power-gate its sparsity-specific components so that performance and energy are not penalized.
\end{itemize}

 We highlight the following experimental observations:

\begin{itemize}
    \item \OURL improves performance by 1.95x on average for data parallel accelerator using processing elements that can perform 16 MAC operations per cycle.
    \item \OURL improves energy efficiency by 1.6x.
     \item Performance improvements with \OURL remain stable throughout the training process.
     \item Considering only the area for compute, \OURL's overhead is 9\% for tiles with 4x4 16-MAC processing elements implementing FP32 arithmetic.
    \item For bfloat16 units, \OURL's compute area only overhead is 13\%.
\end{itemize}

\section{Background and Motivation} 
\label{sec:motivation}
For clarity we restrict attention to convolutional layers, however, our measurements include all layers.
During training, processing a layer comprises three main convolutions:
\begin{alignat}{5}
O & = & W\ &\star\ & A \label{eq:trainingConv1} \\
G_A & = &\ G_O\ & \star& W \label{eq:trainingConv2} \\
G_W & = &\ G_O\ & \star& A \label{eq:trainingConv3} 
\end{alignat}
Where $W$ is the weights, $A$ is the input activations, $O$ is the output activations, $G_A$ is the activation gradients, $G_O$ is the gradients of the output activations and $G_W$ is the gradients of the weights. The first convolution is done during the forward pass to calculate the output activations of the layer while the next two convolutions are done during the back-propagation pass to calculate the input gradients and the weight gradients respectively. Section~\ref{sec:bkground_training} reviews these operations in more detail. Rhu et al., have demonstrated that the activations of convolutional neural networks exhibit significant sparsity during training and proposed compressing the zeros away when transferring data over the PCI-E during training with graphics processors~\cite{rhu2018compressing}. In this section we corroborate these findings and show what levels of sparsity exist in of the three convolutions. Our goal is to exploit sparsity to accelerate the convolutions by eliminating the corresponding MAC operations.

We found that weights exhibit negligible sparsity during training unless the training method incorporates pruning. However, sparsity of the activations and the output gradients is considerable. Thus, we consider exploiting the sparsity of $A$ and $G_O$ in the first and the second convolutions respectively. For the third convolution we target sparsity in $G_O$ or $A$ whichever is higher. The mechanisms we propose can   exploit sparsity for both $G_O$ and $A$ simultaneously. We leave the evaluation of this option for future work.

\begin{figure}%
                \centering
                \includegraphics[width=0.5\textwidth]{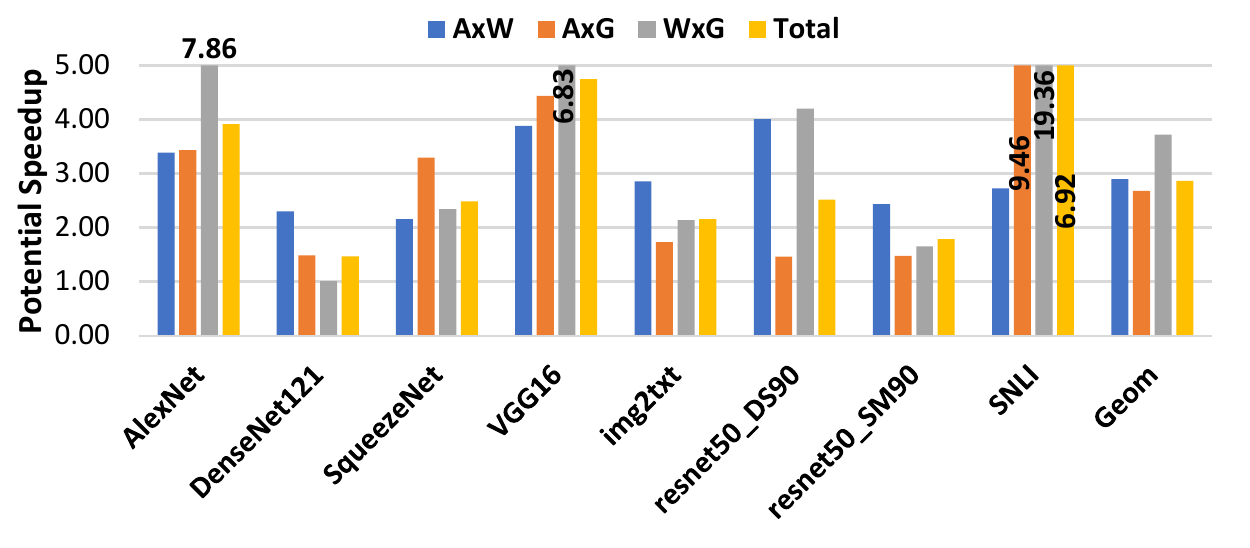}
                \caption{Potential speedup for exploiting dynamic sparsity during training for each of the three convolutions.}
                \label{fig:potentialSpeedup}
\end{figure}

\cref{fig:potentialSpeedup} reports the potential work reduction for each of the three convolutions. The convolutions perform the same number of MACs and take roughly the same amount of time. We report work reduction as a speedup which we define as $\frac{\mathit{all\ MACs}}{\mathit{remaining\ MACs}}$ where $\mathit{remaining\ MACS}$ is the number of MAC operations left are eliminating those where the targeted operand is zero.  On average across all models the potential ``speedup'' for the convolutions is nearly $3{\times}$. The least potential is exhibited by DenseNet121 but even there it is above 50\%. It is more than $2{\times}$ for the highly optimized SqueezeNet. While ResNet50 is a dense network, when trained with two methods that incorporate pruning during training, there is significant sparsity that is induced as the measurements show for resnet50\_DS90 and resnet50\_SM90 (see~\cref{sec:methodology} for the methodology).

\begin{table*}[]
{
\caption{Training Process: Processing of one training sample. Weights are updated per batch (see text).}
\label{tbl:training}
\begin{tabular}{|c|c|}
\hline
\multicolumn{2}{|c|}{\textbf{Forward Pass}}      \\ \hline
\begin{tabular}{m{0.45\linewidth}}
\includegraphics[width=0.45\textwidth]{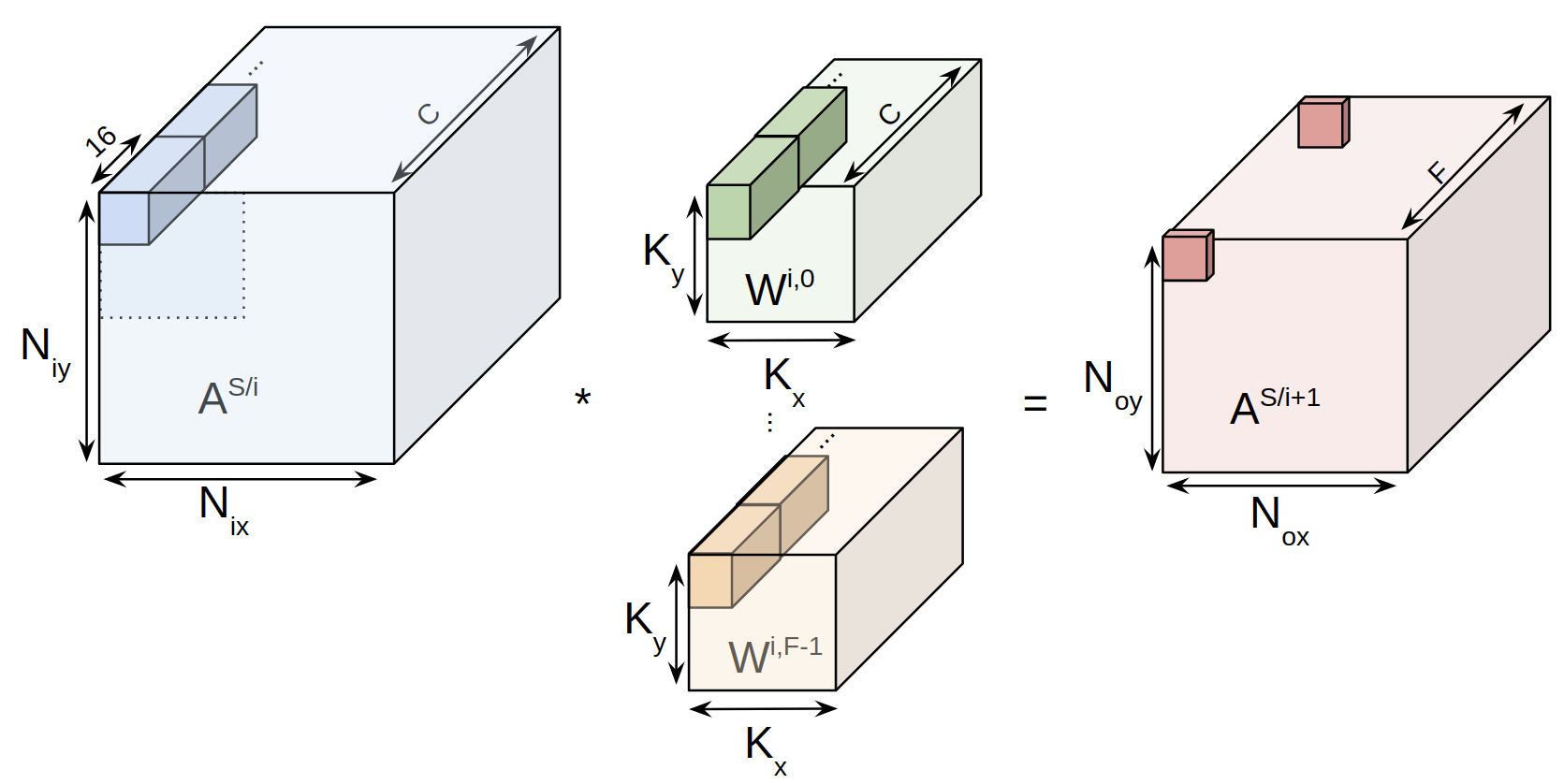}\captionof{figure}{Forward convolution}\label{fig:for_conv}
\end{tabular}
& 
\begin{tabular}{m{0.5\linewidth}}
                    \vspace{.2cm}
                    \textbf{Convolutional Layer: }
                   A sliding-window 3D convolution is performed between the input activations and each of the weight filers to produce one channel in the output activations: 
                   \normalsize{
                   \begin{equation} \label{eq:for_conv}
    A^{S/i+1}_{oc,ox,oy} = \sum^{C}_{ci=0} \sum^{Kx}_{xi=0} \sum^{Ky}_{yi=0} A^{S/i}_{ci,ox+s*xi,oy*s+yi} * W^{i/oc}_{ci,xi,yi}
                   \end{equation}}                    \vspace{-.2cm}  \\  \hline
                    \vspace{.2cm}

                   \textbf{Fully-Connected: } 
                   Each filter produces one output activation:
                   \normalsize{\begin{equation} \label{eq:for_fc}
    A^{S/i+1}_{oc} = \sum^{C}_{ci=0} A^{S/i}_{ci} * W^{i,oc}_{ci}
                    \end{equation}}                      \\ 
\end{tabular}\\ \hline\hline
\multicolumn{2}{|c|}{\textbf{Backward Pass}}     \\ \hline
\multicolumn{2}{|c|}{\textit{Input Gradients}}            \\ \hline
\begin{tabular}{m{0.45\linewidth}}
\centering\includegraphics[width=0.45\textwidth]{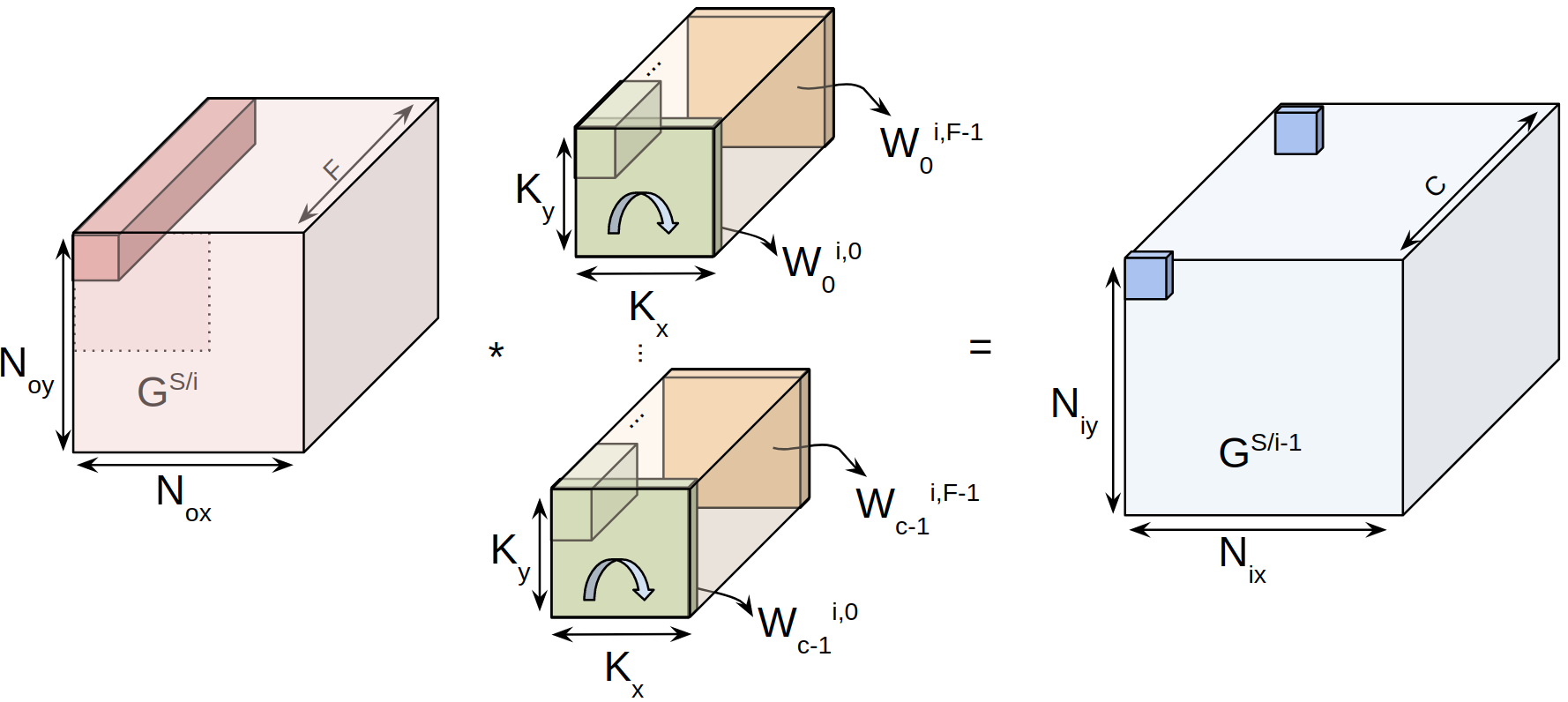}\captionof{figure}{Calculating input gradients}\label{fig:inGrad_conv}
\end{tabular}
& 
\begin{tabular}{m{0.5\linewidth}}
                    \vspace{.2cm}

                    \textbf{Convolutional Layer: }
                   A sliding-window 3D convolution is performed between a reshaped version of the filters with the activation gradients from the subsequent layer. The filters are reconstructed channel-wise and rotated by 180 degrees and the activation gradients are dilated by the stride.                  
                   \normalsize{\begin{equation} \label{eq:inGrad_conv}
    G^{S/i-1}_{oc,ox,oy} = \sum^{F}_{ci=0} \sum^{Kx}_{xi=0} \sum^{Ky}_{yi=0} G^{S/i}_{ci,ox+xi,oy+yi} * {W_{rotated}}^{i,ci}_{oc,xi,yi}
                   \end{equation}}      \vspace{-.2cm}                \\  \hline
                    \vspace{.2cm}

                   \textbf{Fully-Connected: } 
                   The filters are reconstructed and rotated as above. No dilation of the activation gradients.                  
                   \normalsize{\begin{equation} \label{eq:inGrad_fc}
    G^{S/i-1}_{oc} = \sum^{C}_{ci=0} G^{S/i}_{ci} * W^{i,ci}_{oc}
                   \end{equation}}                      \\ 
\end{tabular}\\ \hline
\multicolumn{2}{|c|}{\textit{Weight Gradients}}  \\ \hline
\begin{tabular}{m{0.45\linewidth}}
\centering\includegraphics[width=0.45\textwidth]{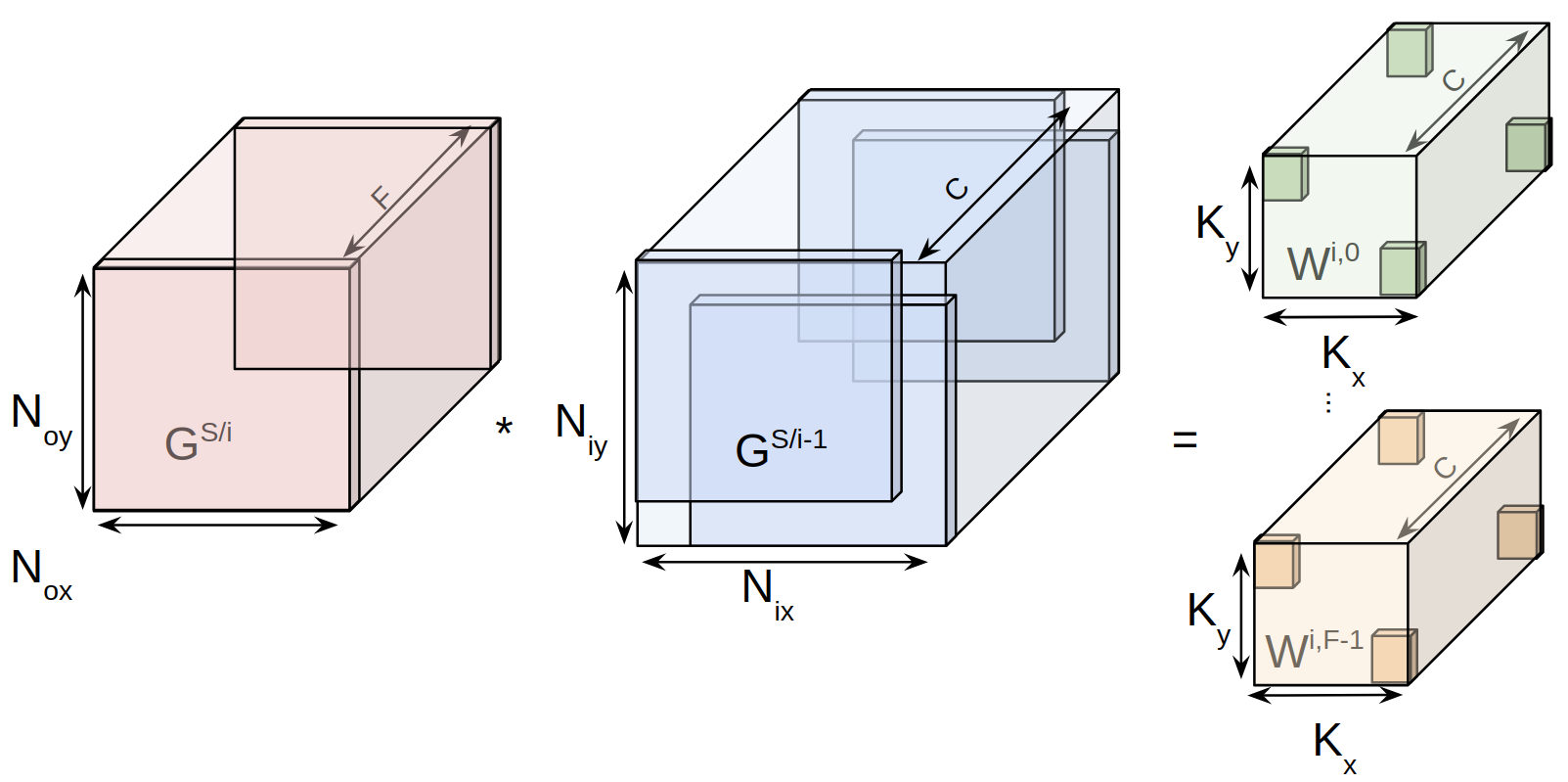}\captionof{figure}{Calculating weight gradients}\label{fig:back_conv_weightGrad}\end{tabular}
& 
\begin{tabular}{m{0.5\linewidth}}
                    \vspace{.2cm}

                    \textbf{Convolutional Layer: }
                   The weight gradients are calculated as a 2D convolution between the input activation of each training sample with its corresponding output gradients which are dilated according to the stride.                  
                   \normalsize{\begin{equation} \label{eq:weightGrad_conv}
    Gw^{i,f}_{oc,ox,oy} = \sum^{S}_{si=0} \sum^{Nox}_{xi=0} \sum^{Noy}_{yi=0} G^{si/i}_{f,xi,yi} * A^{si/i}_{oc,ox+xi,oy+yi}
                   \end{equation}}        \vspace{-.2cm}              \\  \hline
                    \vspace{.2cm}
                   \textbf{Fully-Connected: }
                   Each weight gradient is a scalar product of the input activation and the output activation it affects                  
                   \normalsize{\begin{equation} \label{eq:weightGrad_fc}
    Gw^{i,f}_{oc} = G^{S/i}_{f} * A^{S/i}_{oc}
                   \end{equation}}                      \\ 
\end{tabular}\\ \hline
\end{tabular}
}\vspace{-10pt}
\end{table*}

\subsection{Training Basics}
\label{sec:training_overview}

\label{sec:bkground_training}
\begin{figure}                
\centering
\includegraphics[scale=0.2]{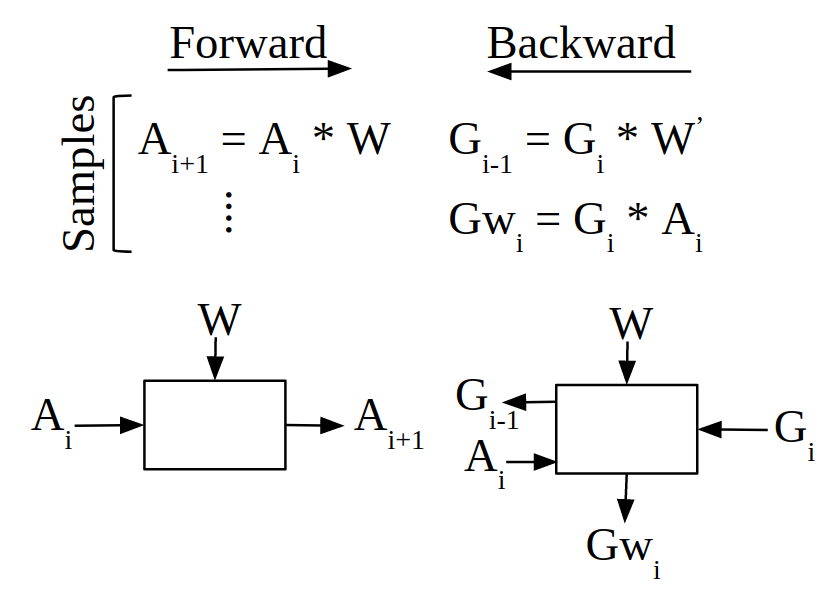}
\caption{Computations during forward and backward phases of training}
\label{fig:training_summary}
\end{figure}

Deep neural networks are trained using a variant of the gradient descent algorithm, where training samples are run through the network to find the prediction error (gradients) relative to the corresponding labels (forward pass) and then to back-propagate these gradients back through the network layers to update the network parameters (backward pass). \cref{fig:training_summary} summarizes the 3 major computations performed per each layer in the network for all training samples> Each computation performs a roughly equal number of operations. We will refer to activations, weights, activation gradients, weight gradients as $A^{S/L}_{c,x,y}, W^{L,F}_{c,x,y}, G^{S/L}_{c,x,y}, Gw^{S/L,F}_{c,x,y}$, respectively where $S$ refer to the training sample, $L$ refers to the network layer, $F$ is the weight filter, $c$ is the channel number, and $x,y$ are the 2D spatial coordinates.

Referring to the three operations shown in Section~\ref{sec:motivation}: During the forward pass, the first operation is applied in sequence from the first to the last layer. At every layer it convolves the weights with the activations to produce the activations for the next layer. Eventually this results into producing the activations for the final layer. These output activations are compared with the known outputs to generate the input gradients for the last layer which will then be back-propagated to update the weights throughout. During back-propagation the layers are invoked in reverse order from the last to the first. Each layer convolves its input gradients with the weights to produce the input gradients for the preceding layer. The layer also convolves the input gradients with the activations to calculate the weight gradients for the layer (the updates for the weights).

The per layer weight gradients are accumulated across the training samples within a mini-batch and used to update the weights once per mini-batch as described by Equation~\cref{eq:weight_update}, where $i$ is the number of weights, $t$ is the epoch number, $\alpha$ is the learning rate, and $S$ is the mini-batch size.
\vspace{-10pt}
\begin{equation} \label{eq:weight_update}
    W^{i}_{t+1} = W^{i}_{t} - \alpha * \sum^{S}_{s=0} G_{s}/S
\end{equation}

\cref{tbl:training} describes the operations in more detail for both convolutional and fully connected layers. For clarity Figures~\ref{fig:for_conv} through~\ref{fig:back_conv_weightGrad} show the operations only for the convolutional layers. A fully-connected layer can be treated as a special-case convolutional layer where all input tensors are of equal size.

\section{Exploiting Sparsity During Training vs. Inference}
\label{sec:sparsity}
\begin{figure}
    \centering
    \includegraphics[width=0.3\textwidth]{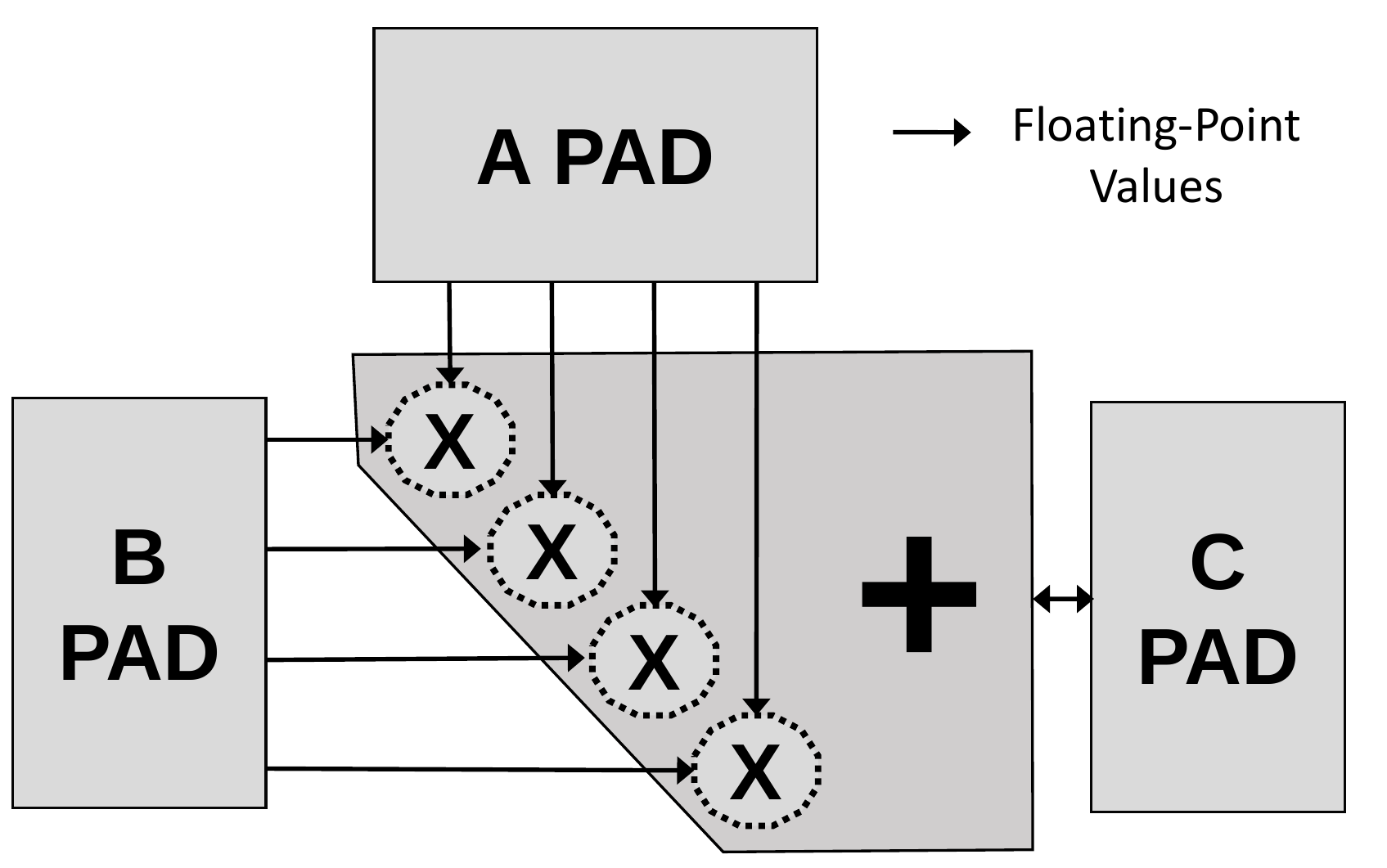}
    \caption{Example Baseline Processing Element.}
    \label{fig:PE}
\end{figure}

For clarity we assume the baseline processing element (PE) shown in~\cref{fig:PE} which can be used as the building block for composing a training accelerator. The PE can perform $N$ (4 in the figure) MAC single-precision \textit{floating-point} operations concurrently all contributing to the same output. For example, these could be $N$ (activation, weight) pairs all contributing to the same output activation. Or they could be $N$ (gradient, weight) pairs all contributing to the same activation gradient. Such processing elements are more energy efficient vs. a single MAC unit because they amortize the energy cost of updating the accumulator over several operations, and the cost of the summation stage by fusing the MACs. The processing element has three local scratchpads, two for inputs and one for outputs. An accelerator may use a grid of these PEs each with separate scratchpads or it may organize several of them in a grid sharing the buffers to exploit temporal and spatial reuse. While we assume single-precision floating point values, \OURL is datatype agnostic and will work with any datatype such as for example bfloat16~\cite{bfloat16}, fixed-point or specialized narrow floating-point~\cite{narrowFP_ibm}. \OURL eliminates MAC operations were at least one of the operands is zero.

\begin{figure*}%
                \centering
                \subfloat[Input Tensors]{
                        \includegraphics[width=0.16\textwidth]{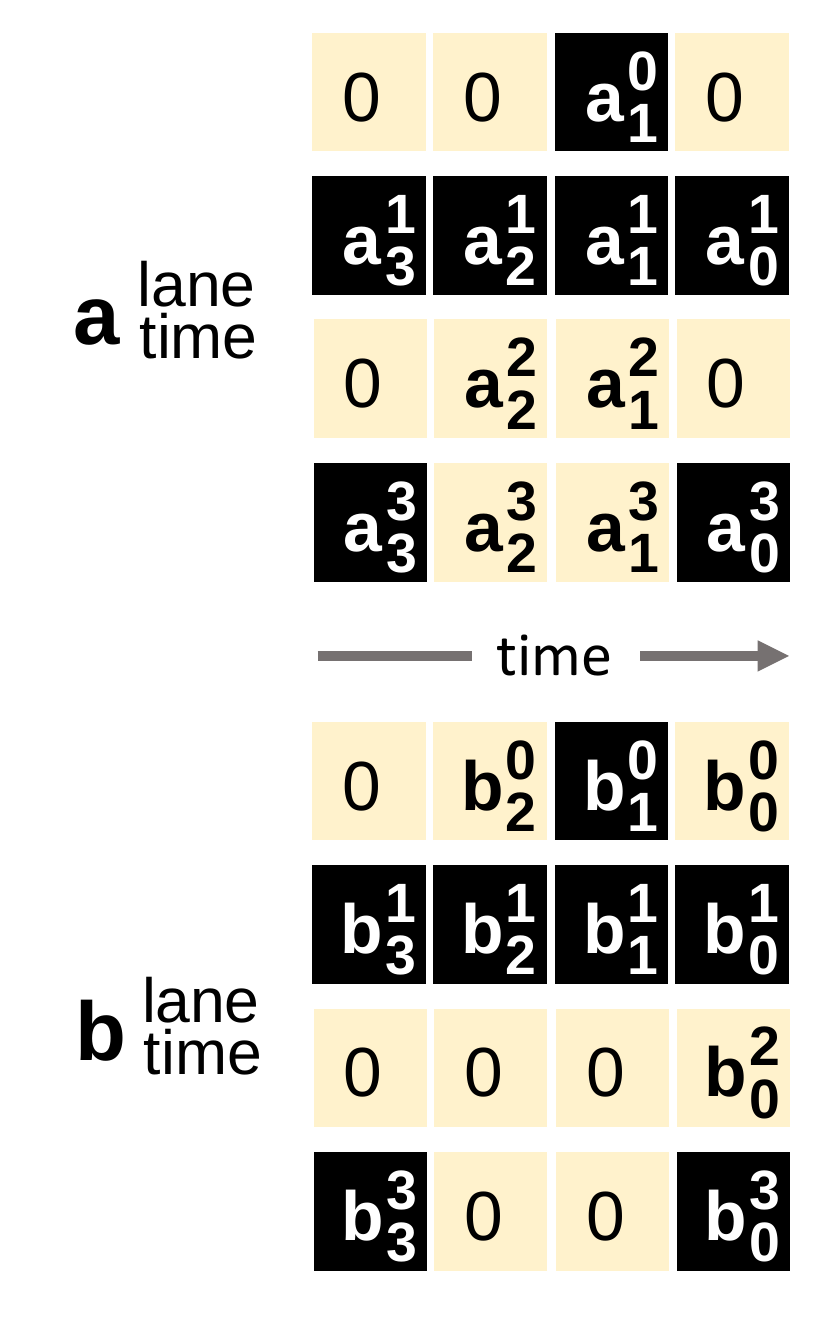}
                        \label{fig:ex:base}
                }\hspace{15pt}
                \subfloat[Unrestricted\newline Movement]{
                        \includegraphics[width=0.10\textwidth]{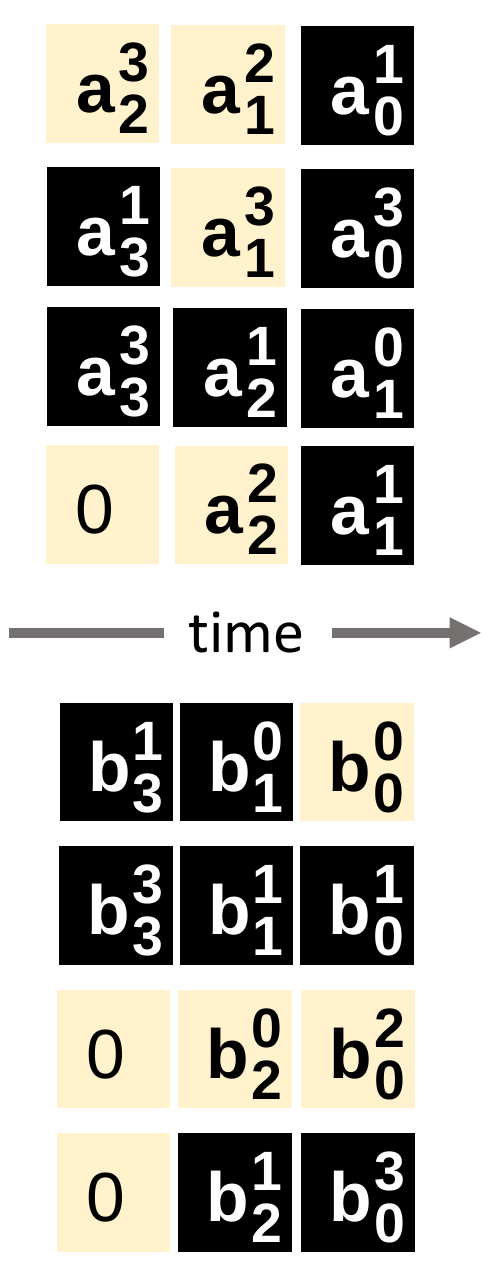}
                        \label{fig:ex:tight}
                }
                \subfloat[Sparse Interconnect]{
                        \includegraphics[width=0.2\textwidth]{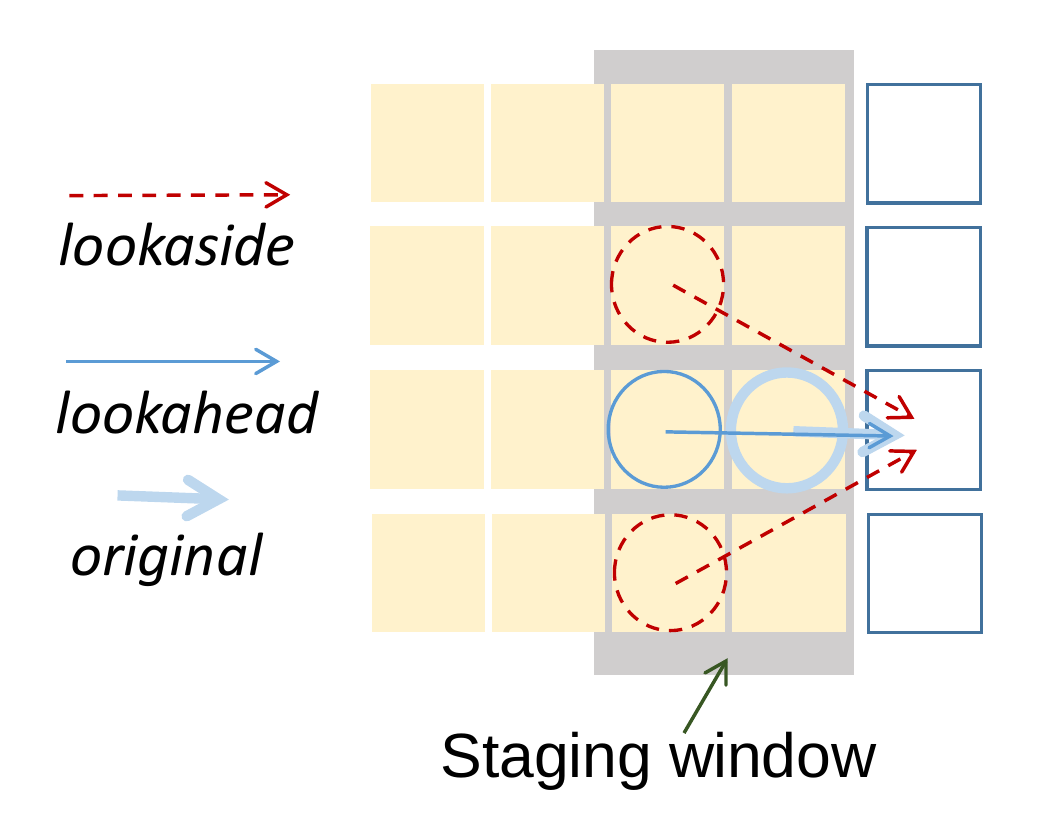}
                        \label{fig:ex:ours}
                }
                \subfloat[Cycle 1]{
                        \includegraphics[width=0.17\textwidth]{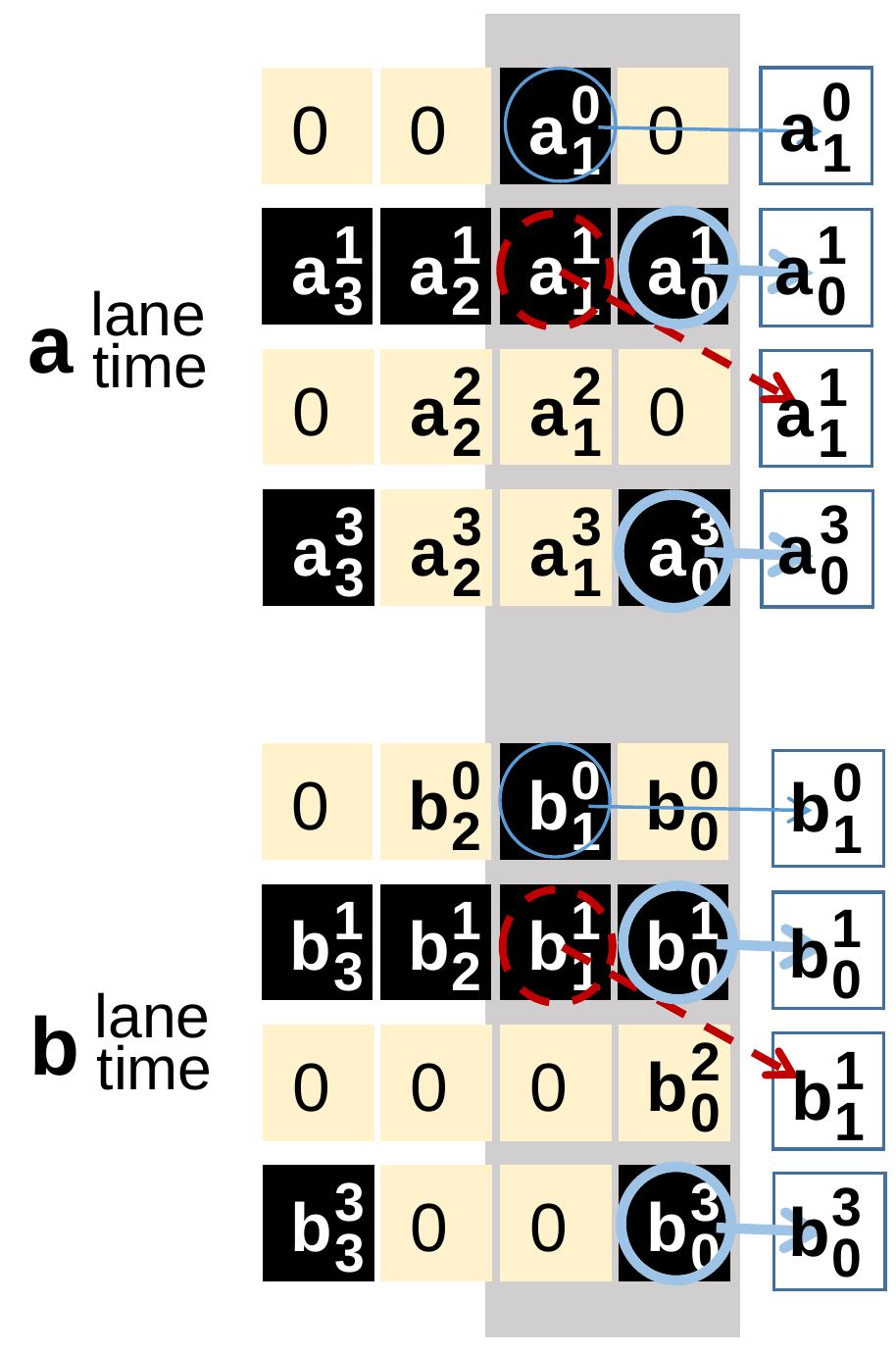}
                        \label{fig:ex:c1}
                }
                \subfloat[Cycle 2]{
                        \includegraphics[width=0.175\textwidth]{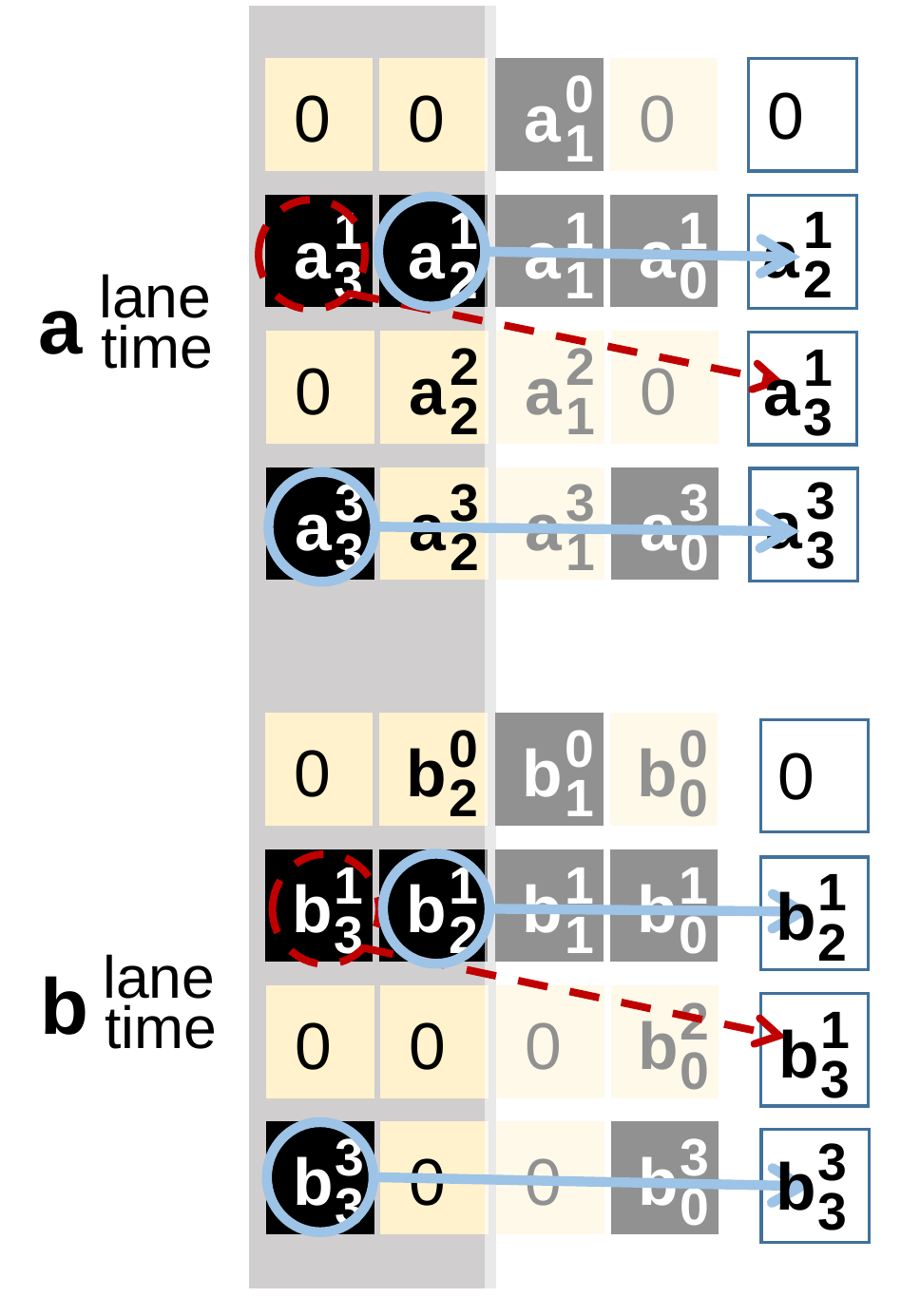}
                        \label{fig:ex:c2}
                }
                \caption{Example of exploiting sparsity dynamically. Allowing a restricted set of movements per multiplier is sufficient.}
                \label{fig:example}
      \end{figure*}

Let us refer to the two input streams as $A$ and $B$ while using $C$ to refer to the outputs. Figure~\ref{fig:ex:base} shows an example of how 16 value pairs will be processed when we do \textit{not} attempt to eliminate those that are \textit{ineffectual} (at least one of the two input values is zero). We denote the input values as $a^{lane}_{time}$ and $b^{lane}_{time}$, where $lane$ designates the multiplier they appear at, and $time$ is the processing order. The figure shows that with the \textit{dense schedule}, that is when we process all values pairs regardless of their value, it is straightforward to arrange them in memory so that the PE can read them as rows from the input buffers performing 4 MACs per cycle. The PE needs 4 cycles to process them.

In the example, however, there are only 7 value pairs (highlighted in black) where both operands are non-zero. As long as the PE processes these value pairs, the output will be correct. The baseline PE of~\cref{fig:ex:base} could take advantage of the \textit{ineffectual} pairs to reduce energy by power-gating the multiplier and part of the adder tree when encountering any of them. For example, Eyeriss used this approach during inference with fixed-point arithmetic~\cite{eyeriss-jssc}. To improve performance and to further reduce energy, \OURL's goal is to eliminate the ineffectual pairs by filling their positions with effectual pairs. Ideally, our 4 MACs/cycle PE should be able to process all effectual pairs in 2 cycles. However, this requires  moving values in tandem from both sides in \textit{time} (earlier yet to the same multiplier) and in \textit{space-time} (earlier and to a different multiplier). 

To exploit sparsity we can draw from the experience with past designs that did so for \textit{inference} alone, e.g.,~\cite{cambricon:2016,CambriconX,SCNN,CambriconS:MICRO2018,Tactical}. Inference executes only the $A{\star}W$ convolution where the weights are known \textit{a priori} and so is their sparsity pattern. Finally, since there is only one convolution and one pass, a single dataflow is sufficient so that we can arrange values in memory in the order we wish to process them. However, for convolutional layers there are multiple windows, which means that weights will have to be matched with different activations per window.  \cref{fig:ex:tight} shows an approach representative of several past designs where the non-zero values from both sides were allowed to \textit{independently} move with no restriction both in time and space-time~\cite{cambricon:2016,CambriconX}. The non-zero values in $A$ are now tightly packed one after the other in memory space and so are the values in $B$. The values belonging to the same pair are no longer aligned in time nor in space. To avoid processing all ineffectual pairs, we need to somehow identify those pairs where both values are non-zero, make them meet at some multiplier. We would also like to keep as many multipliers busy as possible. 
This is a challenging task for two reasons: 1)~Performing arbitrary movement of values in time and space is expensive in hardware. 2)~To keep the 4 multiplier lanes busy, we will often need to grab values from multiple rows from each buffer. In our example, from the first rows of $A$ and $B$ there are only two effectual pairs since $a_0^0$ and $a_0^2$ are zero rendering their corresponding $b_0^0$ and $b_0^2$ ineffectual.

Cambricon is representative of a class designs that exploit sparsity only on the weight side~\cite{cambricon:2016}. Cambricon tightly packs the non-zero weights in memory space so that at run-time the PE can access them a row a time. Each weight is annotated with metadata so that Cambricon can determine which its dense $({lane},{time})$ position. A unit maintaining a pool of activation candidates is tasked with locating and pairing each non-zero weight with its activation. This unit proves expensive as it performs the function of a crossbar so that activations can mirror the arbitrary movement of weights in memory space. Cambricon-X exploits sparsity on both sides allowing weights and activations to freely move both in time and space-time. An indexing module is tasked with matching non-zero weights and activations~\cite{CambriconXMICRO16}. Cambricon-S improves efficiency by imposing \textit{structural} constraints on how the model is pruned~\cite{CambriconS:MICRO2018}. Effectively, it eliminates ineffectual pairs only if 16 of them appear together in a single row. These structural constraints must be imposed during pruning. Cnvlutin2~\cite{Cnvlutin2} and SparTen~\cite{SparTen} exploit sparsity on both sides albeit by paying the cost to deploy independent buffer banks per multiplier input (both sides). They support movement of values only in time and hence cannot effectively handle work imbalance across lanes. ``Struggler'' lanes become a bottleneck. SCNN tightly packs non-zero weights and activations in memory and processes only effectual pairs at runtime. To do so, it processes values one channel at a time so that the product of any weight with any activation is guaranteed to contribute to an output activation. SCNN avoids all data movement at the input. However, it does require a crossbar to route products to accumulator banks. The crossbar is over-provisioned to avoid stalls due to bank conflicts which would otherwise be significant. Bit-Tactical uses a low-cost sparse interconnect at the front-end and a software scheduler to extract sparsity in the weights of pruned models without imposing any restrictions on how sparsity is structured~\cite{Tactical}. On the activation side it targets sparsity within values (bit-level sparsity) and for that it uses shift-and-add multiplier-based MAC units. 

None of the above approaches have been applied in training. We highlight the following differences:  1)~The sparsity pattern during training is always \textit{dynamic}. During inference the weights are statically known and as a result the weights can be easily pre-packaged in memory. 2)~During training, all tensors participate in two convolutions each. The group of values that contribute to an output in each convolution is different and so must be the order in which they are arranged. For example, the filter channels during the forward pass are different from those of the ``reconstructed'' filters during the backward pass  (The ``reconstructed'' filters during the backward pass are formed by taking the weights from the same channel across all filters, stacking those along the channel dimension and then transposing the filter). Similarly, the gradients need to be bundled together differently for the second convolution and the third. These are calculated per layer during the backward pass where we would like to avoid having to spill the gradients off-chip. There is no single way to pack them in memory (effectively pre-scheduling them) that would work for all cases where they are used.  3)~Activations can be discarded after each layer during inference which is not the case during training. 4)~Inference accelerators used narrow fixed-point arithmetic. Training today is done predominantly using floating-point. Floating-point values are wider making crossbars considerably more expensive than narrow fixed-point data, and performing shift-and-add operations is non-trivial for floating point.

In this work we borrow upon the sparse-interconnect/limited-movement-options approach used by Bit-Tactical's front-end and adapt it so that it can be used during training. In particular, we wish to use a low-cost sparse interconnect to \textit{dynamically} eliminate ineffectual value pairs at runtime. However, compared to Bit-Tactical there are the following major differences and challenges: 1)~While Bit-Tactical used a software scheduler for packing weights in memory, the dynamic nature of sparsity during training makes this approach impractical. The overhead of invoking a software scheduler per layer/sample/convolution is prohibitive in terms of latency and energy. 2)~Bit-Tactical pre-schedules values (weights) packing them in memory in bundles so that they can be fetched and processed together. This is possible during inference since  the weights are being used only in the first convolution above, and where weights and activations are accessed in one specific order. Unfortunately, during training this is no longer possible. Each tensor is accessed in two different orders across the three convolutions. 3)~Bit-Tactical used fixed-point shift-and-add units. Training in general requires floating-point units. 

\begin{figure}
    \centering
    \includegraphics[width=0.3\textwidth]{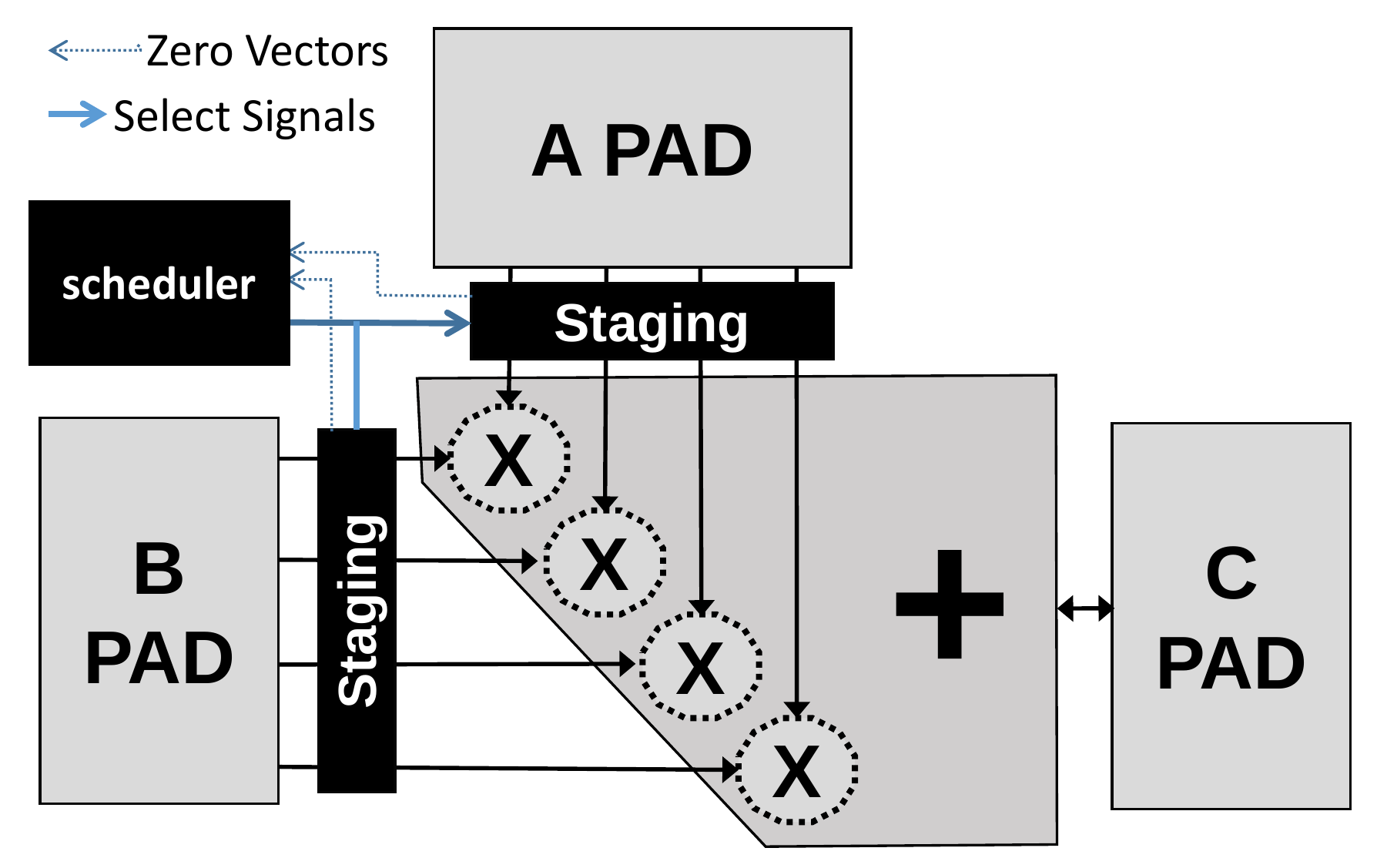}
    \caption{\OURL Processing Element.}
    \label{fig:ourPE}
\end{figure}

\subsection{\OURLCORE}
\label{sec:architecture}
Here's how \OURL removes ineffectual values pairs when processing the example input tensors of Figure~\ref{fig:example}. Let us assume that we are processing the 3D convolution of two input tensors $A$ and $B$ and for clarity let us assume that our processing elements perform 4 MAC operations concurrently.  

Figure~\ref{fig:ourPE} shows that the \OURL PE extends the baseline PE with the following components: a)~There is now a staging buffer for $A$ and another for $B$. Each staging buffer can hold up to two rows. Writes to these stage buffers are row-wide. There are 4 reads ports each feeding directly to a multiplier input. The connectivity per read port is sparse: each port can read out one out of a limited set of values (4 in our example) within the staging buffer. The set of values that each port can read out is different but can overlap. b)~There is a hardware scheduler. The hardware scheduler accepts a bit vector from each staging buffer identifying which values are non-zero. For 2-deep staging buffers, the bit vectors would be 8b wide for our example. Each cycle the scheduler selects up to 4 effectual pairs from the staging buffers. It generates the control signals for the read ports (2b per port for our example) so that the corresponding values are read out. The same control signal is shared among the corresponding ports in the two staging buffers, i.e., the same control signal goes to port $p$ in the horizontal and vertical staging buffers so both operands move in tandem (4x2b control signals in total).
 
The example of Figure~\ref{fig:ex:ours} shows that, per read port, \OURL allows only a limited set of value movements per multiplier. There are two types of movement: in \textit{time} only or \textit{lookahead}, and in \textit{space-time} or \textit{lookaside}. The figure shows the set of movements for second multiplier: it can either process the original dense value $a^1_0$, the next value in the same lane $a^1_1$ (lookahead), or it can \textit{steal} the values from a step ahead in time from its two neighboring lanes $a^0_1$ or $a^2_1$ (lookaside). In our example, the movements possible by the other read ports are structurally identical relatively to their lane (the ports are treated as if they are arranged into a ring with port 0 being adjacent to port 3). However, each port can access a different set of values.
Figures~\ref{fig:ex:c1} and~\ref{fig:ex:c2} show how \OURL reduces processing time to the minimum 2 cycles using just a 4-input multiplexer per multiplier input. 

To improve performance, the staging buffers will need to be kept full with values as much as possible. Accordingly, the $A$ and $B$ buffers will have to be banked accordingly to sustain a higher read throughput. For our example two banks would be sufficient. In general, we would like to have at least as many banks as $lookahead$. We have found empirically that a lookahead of 3 is more than sufficient. We described our preferred PE configuration and the hardware scheduler next.

\subsection{The Hardware Scheduler}\label{sec:scheduler}
\begin{figure}%
                \centering
                \includegraphics[scale=0.33]{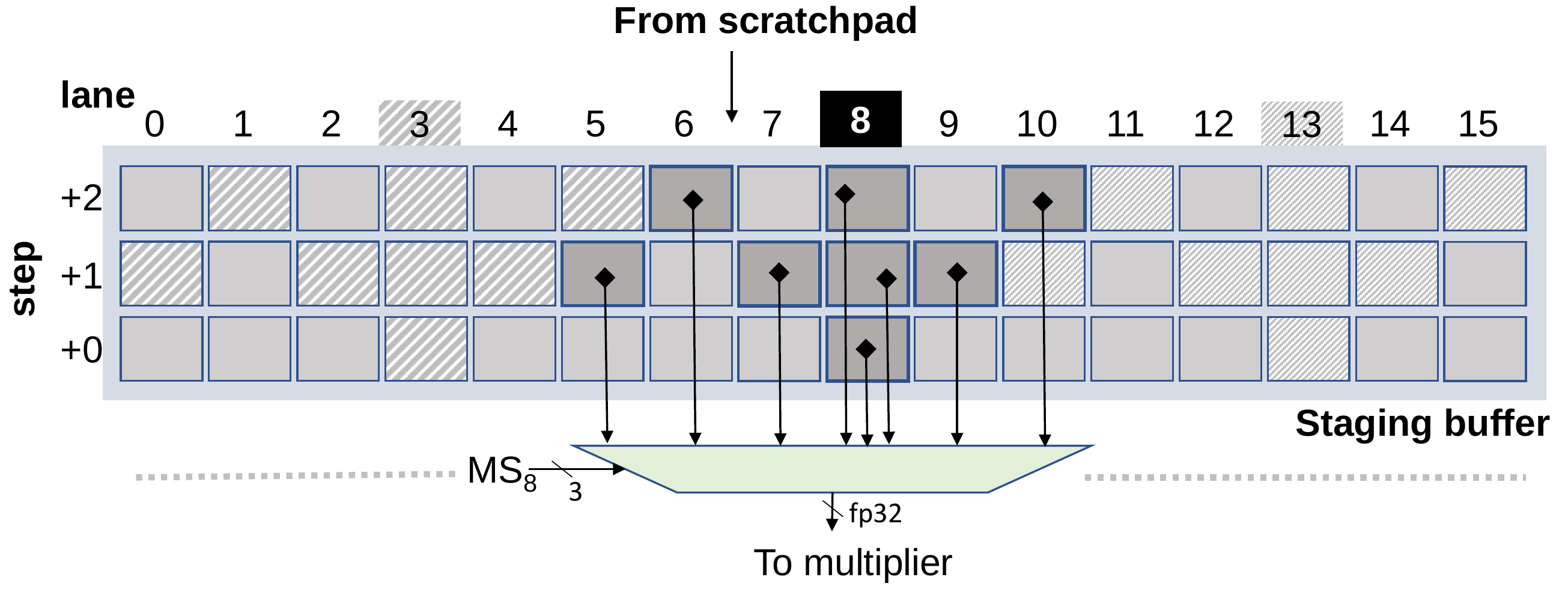}
                \caption{Staging buffer connectivity for the 16-input MAC \OURL PE. Shown is the connectivity for lane \#8.}
                \label{fig:promo_map}
\end{figure}

Our preferred PE processes 16 MACs per cycle. It accepts 16 pairs of $(A,B)$ single-precision floating-point values. Each input side has a 3-deep staging buffer. Figure~\ref{fig:promo_map} shows one of the staging buffers. Each of the 3 rows contains 16 values corresponding to the dense schedule for the current step (step +0), and the next two in time (+1 and +2).  For every lane there is a multiplexer which implements a sparse connectivity pattern. The figure shows the connections for lane 8.  Besides the original ``dense'' schedule value, there are 2 lookahead and 5 lookaside options per input. For example, the multiplier for lane \#8 can be given the value at lane 8 from the current time slot or up to 2 ahead. Alternatively, it can ``steal'' the values from neighboring lanes. For example, it get the value from lane 6 that is 2 time steps ahead or the value from lane 5 that is 1 step ahead. Each lane has the same connectivity pattern which is shifted relative to its position (wrapping around the ends). This connectivity pattern per input has been shown to work well when extracting sparsity during inference~\cite{Tactical}. The staging buffer also generates a 3x16b zero bit vector indicating which of the values are zero.  The staging buffer has three write ports one per row.

\begin{figure}%
                \centering
                \includegraphics[scale=0.4]{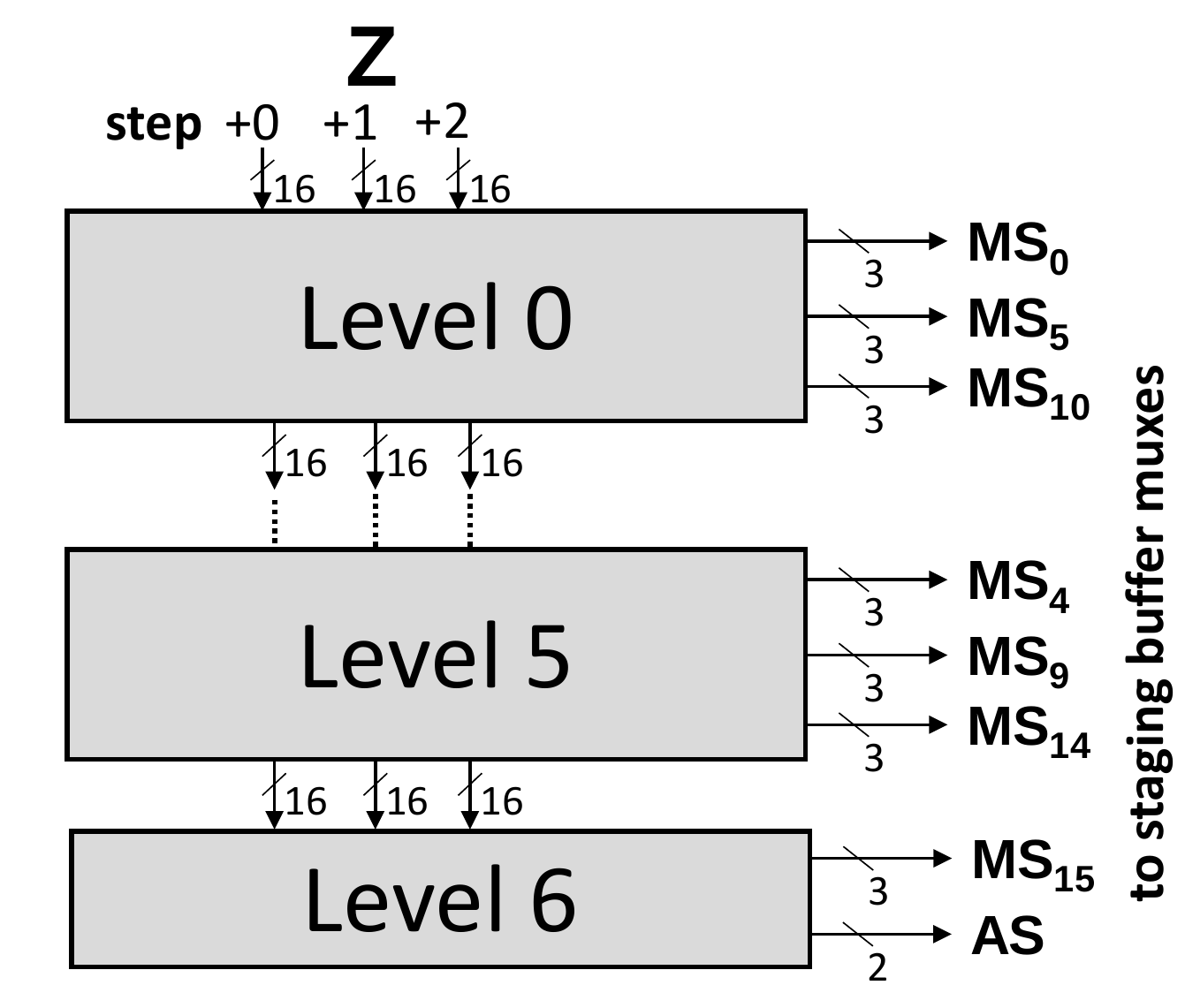}
                \caption{\OURL's Scheduler.}
                \label{fig:scheduler}
\end{figure}

The scheduler accepts the two zero bit vectors $AZ$ and $BZ$ from the $A$ and $B$ staging buffers and generates two sets of signals. The first set is for 16 $MS_i,i{=}0...15$ 3b signals one per input lane. These are the select signals for the per lane multiplexers. There is one $MS_i$ signal per multiplier and it used by the multiplexers on both the $A$ and $B$ sides for the lane. The scheduler also produces a 2b $AS$ signal that indicates how many rows of the staging buffer it has been able to drain so that they can be replenished from the scratchpads (which are banked so that three rows to be read per cycle if needed). 

The rest of this section describes the scheduler block. The $AZ$ and $BZ$ 3x16b bit vectors are first ANDed together bitwise to produce a single $Z$ 3x16b bit vector. This indicates which pairs of $(A,B)$ values have at least one value that is zero. These pairs are ineffectual and can be skipped. The goal of the scheduler is to select a movement per lane, for a total of 16 movements ($MS_i$ signals) so that it uses as many of the remaining $(A,B)$ pairs as possible in one step. We will refer to the selection of movements that the scheduler makes for one step as a \textit{schedule}. 

For each lane $i$ the scheduler uses a simple, static priority scheme: among the 8 options select the first available in the following order (notation is (step,lane) refer to Fig.~\ref{fig:promo_map}): (+0,$i$) (dense schedule), (+1,$i$) lookahead 1 step, (+2,$i$) lookahead 2 steps, and then the lookaside options: (+1,$i$-1), (+1,$i$+1), (+2,$i$-2), (+2,$i$+2), and (+1,$i$-3). A 8b-to-3b priority encoder suffices. However, having all lanes make their selections independently may yield an invalid schedule; the same pair may be chosen by multiple lanes and end up been used more than once. 

To ensure that the scheduler always produces a valid schedule (one where each value pair is selected \textit{once}) we use a  hierarchical scheme where scheduling in done in 6 levels as shown in Fig.~\ref{fig:scheduler}. In each level, a subset of the lanes make their decisions independently using the current value of the $Z$ vector as input. The lanes assigned at each level are guaranteed by design to not being able to make overlapping choices. After they make their selections they ``remove'' these options (AND gates) from the $Z$ vector before passing it to the next level. Figure~\ref{fig:promo_map} shows that the options for lanes \#3, \#8, and \#13 are non-overlapping \textit{by design}. Following a similar reasoning we can arrange all priority encoders into 6 levels, with 3 lanes per level for the first 5 levels and 1 lane for the last. The lane groups per level are: \{0,5,10\}, \{1,6,11\}, \{2,7,12\}, \{3,8,13\}, \{4,9,14\}, and \{15\}. Generating the AS signal is straightforward given the bits that are left enabled in $Z$ at the end.
While we have described the above process in steps, the scheduler is \textit{combinatorial} and operates in a single cycle. 


\subsection{Composing Tiles}\label{sec:tiling}
So far we have described a single \OURL processing element (PE) which can exploit sparsity on both operands. An accelerator can use multiple such PEs to achieve a performance target. This PE can exploit reuse only temporally. To take advantage of data reuse also spatially we can organize multiple PEs in a grid where PEs along the row share the same B input and PEs along the same column share the same A input. For example during the forward pass and for a convolutional layer, each row can be processing a different filter, whereas columns can be processing different windows. In this arrangement each PE would be processing a unique combination of B and A inputs. Skipping zeros on both A and B sides remains possible if we use per PE schedulers and staging buffers. 

\begin{figure}%
                \centering
                \includegraphics[scale=0.4]{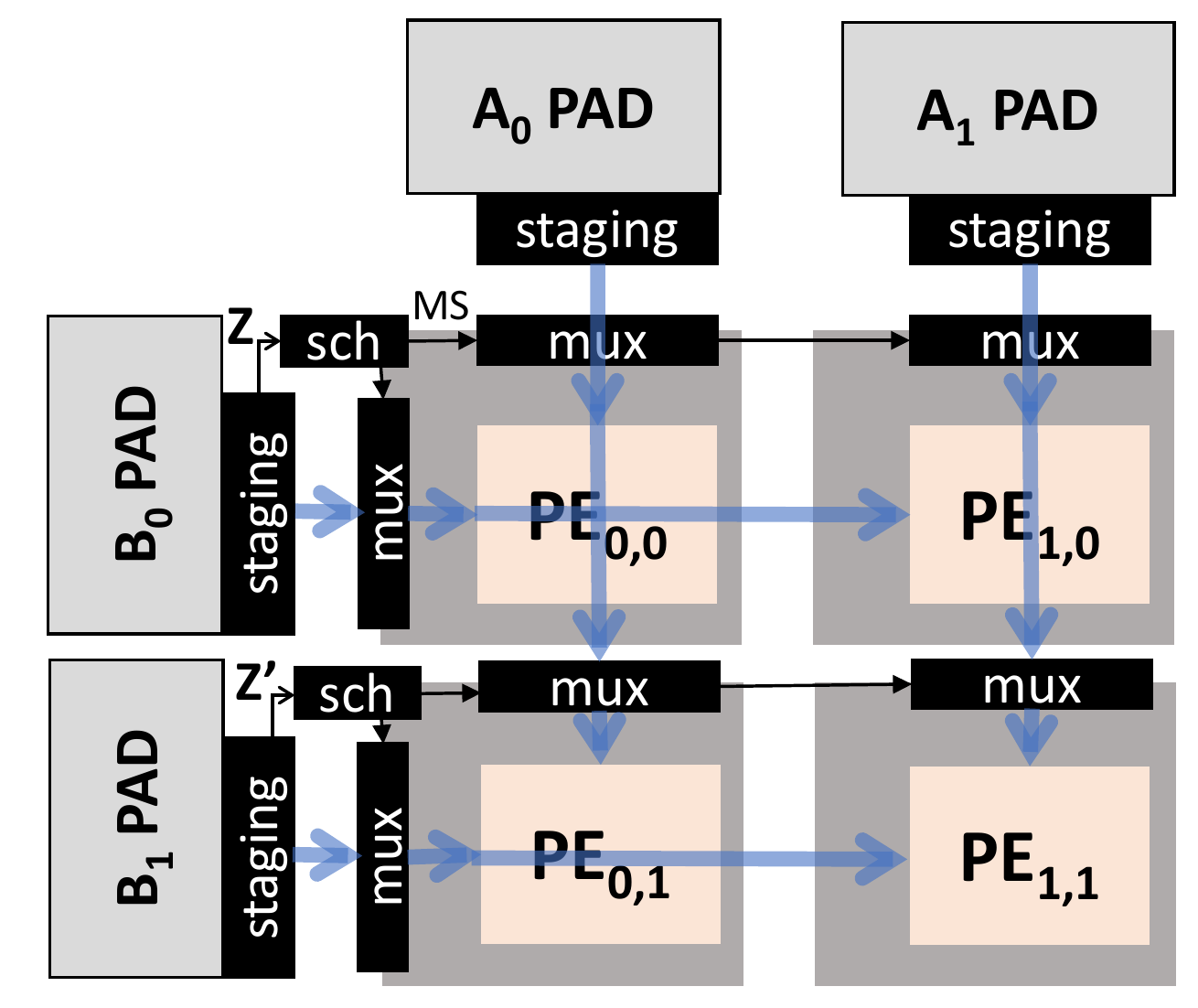}
                \caption{A 2x2 \OURL Tile.}
                \label{fig:tile}
\end{figure}

In the designs we evaluate we do use tiles comprising a grid of multiple PEs. However, we opt for extracting sparsity from only the $B$ side; there is sufficient sparsity on one of the operands in each of the three major operations to extract significant benefits. Figure~\ref{fig:tile} shows an example configuration of such a tile. The tile uses a common scheduler per row and shares the staging buffers for the B side. For the A side, it uses a single staging buffer per column and separate multiplexer blocks per PE. The A-side multiplexer blocks per row share the $MS_i$ from the row scheduler. The schedulers now need to see only the $Z$ vector from their B-side staging buffer.

\subsection{Tensor Layout and Transposing}\label{sec:tensorlayout}
During training, some of the tensors are used in more than one of the major computations. For example, the weights in the forward pass are convolved with the activations whereas in the backward pass are convolved with the output gradients. In each case the group of values that contribute to each output value is different. This has implications for the memory hierarchy which needs to supply the data in appropriate order to the PEs. When a tensor is used in only one way it is possible to statically layout  the values in memory so that they can be easily served using wide accesses off- and on-chip. However, during training the layout that serves well one of the computations will not be able to serve well the other. Fortunately, it is possible to arrange values in memory so that they can be easily fetched for all use cases. The key is the ability to transpose tensors as needed. For this purpose, we use a tensor layout where values are stored in groups of 16x16 values. The group is formed by taking 16 consecutive blocks of values along the row dimension. Each of these blocks contains 16 continuous along the channel dimension values. The starting coordinates for each 16x16 value group are aligned by 16 along the row and the channel dimensions. Finally, the groups for a tensor are allocated in memory space in channel, column, row order. 

When fetching values from off-chip each group can be written directly to the multi-bank on-chip memories so that each 16-value block is copied directly to a bank. As a result, the PE can now directly access any block of 16 consecutive along the channel dimension values in a single step. When transposing is needed, we use on-chip transposers between the on-chip memory banks and the tile scratchpads. The number of transposers used can be chosen so that they memory system can supply data at a sufficient rate to maintain the tiles busy.  Each transposer reads 16 16-value block from their banks using 16-value wide accesses. It copies those into its internal 16x16 buffer. The transposer then can provide a group of 16 values composed of a single value from each of the 16 groups it read from memory effectively transposing the tensor. For example, it can supply all values that appear first within their original block, or all that appeared third. This is needed for the weights and the gradients.

\subsection{Models with no Sparsity}\label{sec:nosparsity}
While many models exhibit sparsity during training not all will. When there is no or little sparsity we would like to avoid hurting performance and energy efficiency. Fortunately this is straightforward by power-gating the \OURL-specific  components and by bypassing the staging buffers. The decision to power-gate can be taken statically if it is known that the model will exhibit no sparsity. Alternatively, as the model is training a counter per tensor at the output of each layer can measure the fraction of zeros that were generated. This information can be used to automatically decide whether enabling \OURL for the next layer would be of benefit. This is possible in the forward and the backward pass. 

\subsection{Keeping Tensors Scheduled In Memory}
Thus far we assumed that the tensors are kept in dense format in memory, which is to say that zeros are also stored. Off- and on-chip we can use any of the memory compression techniques previously proposed (e.g., zero compression via run-length encoding~\cite{han_eie:isca_2016,SCNN}) to keep the tensors in compressed form. However, prior to passing them to \OURL we have to decompress them to the dense form so that \OURL can schedule them for execution. Alternatively, we can use the scheduler of \OURL as a compression engine. In this section we describe several options for doing so.

We can extend \OURL so that it can store both input tensors in scheduled form in memory. In this case, each value is stored as a pair $(v,idx)$ where $v$ is the value and $idx$ is the movement it performed.
The $idx$ is equivalent to the $MS$ signal that the front-end scheduler would have produced given this tensor alone (one-side scheduling). Ideally, only non-zero values will be stored and the scheduling approach of \OURL is used as a memory compression technique. Provided there is sufficient sparsity, this approach reduces footprint and the number of accesses needed to read the tensor. Further, it amplifies on-chip memory capacity and in turn can reduce accesses to higher levels of the memory hierarchy and more importantly to off-chip memories.

\subsubsection{Fully-Connected Layers During Inference }
We describe this approach first only for the weight side of fully-connected layers during inference. We then describe how it can be extended to handle both weights and activations, convolutional layers, and training. During inference, the input tensors  to a fully-connected layer are the activations and several filters (weights). Each filter produces a single output activation by multiplying each input activation with a weight while accumulating the product into the output. In this case, both input tensors are accessed in one specific way and, thus, we can choose a convenient processing order. 

\noindent\textbf{Pre-Scheduling Weights: }To exploit sparsity on the weight side only, we can simply statically pre-schedule the weight tensor for each filter. In this case, we do not need to the use the dynamic scheduler at all and we can bypass the staging buffer on the weight side. The multiplexer signals for the activation-side staging buffer can be directly driven by the $idx$ fields of the weights. The on-chip memory hierarchy must be modified to accommodate these $idx$ fields and provide connection from them to the multiplexers. This is similar to the Tactical front-end software scheduler~\cite{Tactical}. 

\noindent\textbf{Pre-Scheduling Activations: }Since activations are generated at runtime as an output from the preceding layer we have to schedule them at runtime.  Fortunately, this can be achieved by implementing a back-side scheduler which operates at the output of the PEs. This is described in Section~\ref{sec:bsched}.

\begin{figure}%
                \centering
                \includegraphics[scale=0.4]{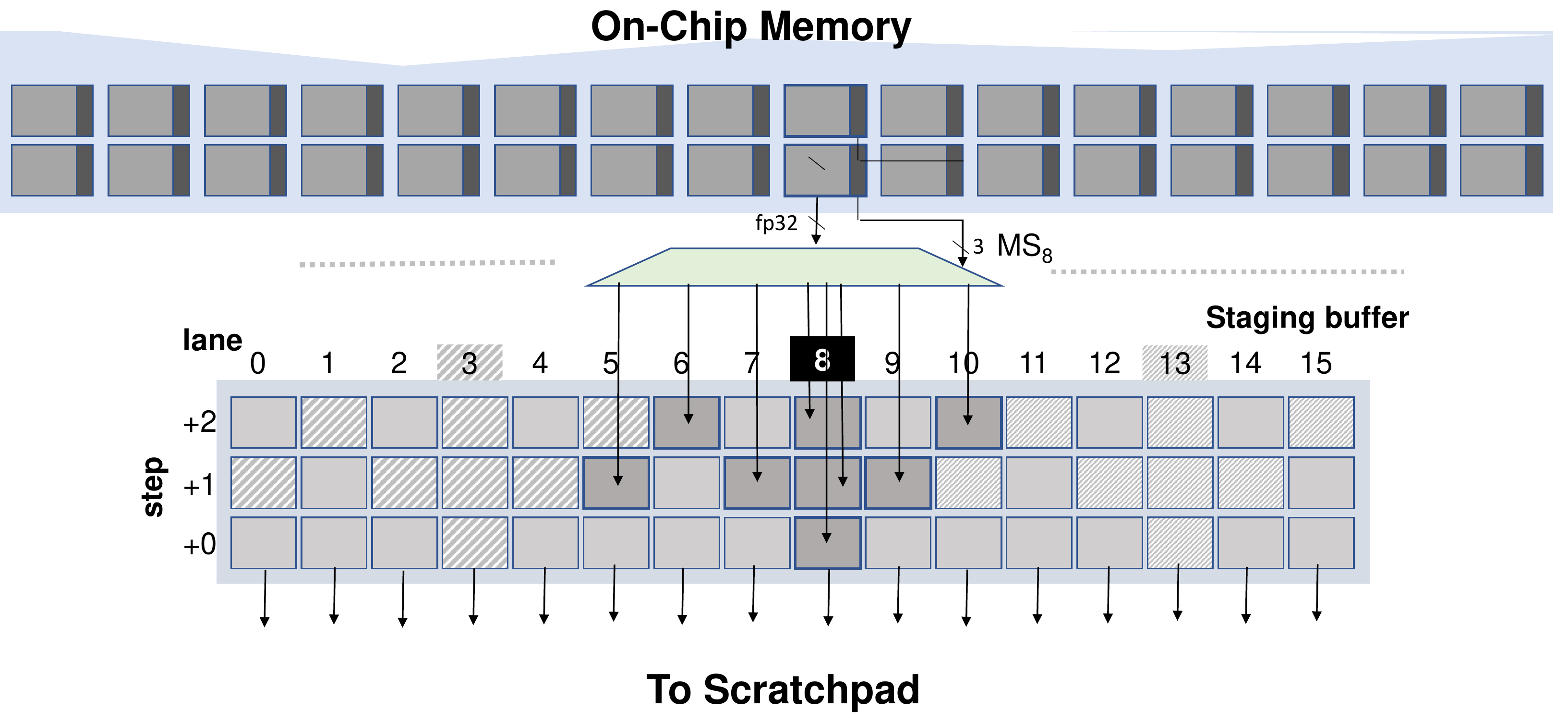}
                \caption{Decompressing a Scheduled Tensor to its Dense form. Shown is the decompressing logic for element 8 within a row of 16 elements. The decompression uses the promotion map of Figure~\ref{fig:promo_map}.}
                \label{fig:revmux}
\end{figure}

\noindent\textbf{Pre-Scheduling Both Activations and Weights: }It is also possible to take advantage of sparsity on both sides. Here both tensors are stored in scheduled form in memory. However, prior to copying the tensors to a PE's scratchpads they are expanded to dense form. Figure~\ref{fig:revmux} shows the hardware needed for performing this decompression. Essentially, this is the mirror of the multiplexer stage of the previously described \OURL scheduler. Since the tensors are now in their original dense format in the scratchpads, \OURL can reschedule them to take advantage of sparsity on either or both sides.

\subsubsection{Pre-Scheduling for Convolutional Layers}
There is an additional challenge for convolutional layers. Again let's focus first solely on inference.  When we pre-schedule a tensor, we do so assuming a specific processing order in which the whole tensor will be processed. The values that appear in a single step of this schedule are meant to be processed together by a PE and thus must contribute to the same output value. Given that we consider inference only now, this can be easily handled for the weights regardless of the layer type. In convolutional layers however, each activation participates in several windows. For example, assuming 3x3 filters and a stride of 1, each activation will participate in 9 different windows. Accordingly, there is not a single processing order through the activation tensor that we can use to pre-schedule it. However, regardless of the window, the activations with the same $(\mathit{row},\mathit{column})$ coordinates will always be used together. Accordingly, we can at least schedule activations in groups across the channel dimension. For example, for a layer with 128 channels and for an accelerator with PEs with 8 MACs, we can schedule the activations in groups of 128. All the activations in a group will have the same $(\mathit{x,y})$ coordinates while the channel $c$ takes all possible values for the layer (0 to 127). The dense schedule would require $128/8$ steps giving us able opportunities to reduce the number of steps needed to process the activations per group. The schedule in this case will not be allowed to span across different groups when the stride is one. It may be able to do so for larger strides where some groups, i.e., $(\mathit{x,y})$ coordinates, will never be used as starting point of a window. For example with stride 2, if a window starts at $(\mathit{x,y})$ then there will be no window starting at $(\mathit{x+1,y})$. This means that the schedule is free to span across these two groups effectively treating them as one large group. And given that typically the stride applies to both the $x$ and $y$ coordinates we will be able to schedule together four groups starting respectively at $(\mathit{x,y})$, $(\mathit{x+1,y})$, $(\mathit{x,y+1})$, and $(\mathit{x+1,y+1})$.

To process the layer, however, we need to be able to access the activations that belong to each window. If we use \OURL's scheduling to compress them in on-chip memory, then the location of each of the groups belonging to the window will vary and we will not be able to directly calculate it based solely on its $(\mathit{x,y})$ coordinates. One option would be to keep an additional pointers to each scheduled group. Another is to have each group starting at the memory location it would start at if it is stored in dense form. That is, the group is scheduled and fills up as much space as it needs, however, we reserve for it enough space for the worst case (no sparsity) regardless. In this case, we do not reduce the amount of on-chip memory needed. However, we still benefit from reducing the amount of data that will be read and written on-chip. Accordingly, it will reduce energy consumption of on-chip accesses.

Alternatively, we can group activations for compression with \OURL scheduling in groups of $\mathit{16x16}$ as described in Section~\ref{sec:tensorlayout}. We found this grouping scheme to be convenient for the processing order of both forward and backward passes as well as our compute structures. We can schedule these groups for the purpose of reducing the amount of memory space they occupy in which case we will still need pointers to the beginning of each group. Or, as mentioned above, we can allocate enough memory for the worst case and use scheduling to reduce only the number of accesses and thus energy. The scratchpads will have to be large enough to allow us to read in and expand as many groups as necessary according to the dataflow in use.

\subsubsection{Pre-Scheduling During Training}
As we discussed, during training, all tensors are being used in two different ways. Accordingly, it is not possible to create one schedule that would work for both uses. However, we can compress the tensor using a convenient group as described above. For example, in groups of $\mathit{16x16}$ values and expand those just before writing them to the scratchpads for processing. Again the scratchpads will have to be large enough to accommodate all the groups needed to be accessed concurrently according to the dataflow in use. This is necessary if we want to avoid having to read values multiple times.

\subsection{A Backside Scheduler}
\label{sec:bsched} 
Rather than scheduling the A or B\ input tensors just before the PEs, we can instead position the scheduler on the output of the PEs. Doing so allows us to pre-schedule the output values as they are produced and to store  them in scheduled form in memory. That is, each value is stored as a pair $(v,idx)$ where $v$ is the value and $idx$ is the movement it performed.
The $idx$ is equivalent to the $MS$ signal that the front-end scheduler would have produced given this tensor alone (one-side scheduling).

Using a back-side scheduler has several advantages. First, provided there is sufficient sparsity, storing the values in the scheduled form in memory reduces footprint,  reduces the number of accesses needed to read the pre-scheduled tensor, amplifies on-chip memory capacity and in turn can reduce accesses to higher levels of the memory hierarchy and more importantly to off-chip memories. 

Second, given that for typical layers computing an output value entails several MAC\ operations the back-side scheduler can be \textit{iterative}. An iterative scheduler can reuse only one level of those shown in \cref{fig:scheduler} over several cycles to schedule a block of values. For example, for our preferred 16-MAC\ PE, such a scheduler can take 6 cycles to schedule a block of values with the benefit of being less expensive in terms of hardware overhead.
\ 
\section{Evaluation}\label{sec:eval}
\label{sec:methodology}
\noindent\textbf{DNN models}: We evaluate \OURS on models from a variety of applications:~1) image classification trained on ImageNet~\cite{imagenet}: AlexNet~\cite{alexnet}, DenseNet121~\cite{densenet121}, SqueezeNet~\cite{SqueezeNet}, VGG~\cite{vgg}, ResNet-50~\cite{resnet}, 2)~scene understanding: img2txt~\cite{img2txt}, and 3)~natural language modeling: SNLI trained on the Stanford Natural Language Inference corpus~\cite{snli}. We train two variants of ResNet-50: 1)~resnet50\_DS90: following the method of Hesham~\textit{et al.}~\cite{dynSparse}, and 2)~resnet50\_SM90: following the method of Dettmers~\textit{et al.}~\cite{sparseMom}. The two methods incorporate pruning during the training process. For both techniques we target 90\% sparsity.

\noindent\textbf{Collecting Traces}: We train all models using 32-bit floating point on a latest generation commodity graphics processor unit (GPU). We trained each model for as many epochs as needed for it to converge to its state-of-the-art output accuracy. For each epoch, we sample one randomly selected batch and trace the operands of the three convolutions shown in~\cref{eq:trainingConv1,eq:trainingConv2,eq:trainingConv3}; the filters, the input activations per layer, and the output gradients per layer.  The batch size is different per model due to their different GPU memory requirements. It ranges from as low as 64 and up to 143 samples per batch.

\noindent\textbf{Accelerator Modeling}: We developed a custom cycle-accurate simulator to model performance. \cref{tbl:arch_config} reports the default configurations for all architectures studied. To model area and power consumption all designs were implemented in Verilog and synthesized through the Synopsys Design Compiler~\cite{synopsys_site}. Layout was performed using Cadence Innovus~\cite{innovus} and for a 65nm TSMC\ technology (which is the best that is available to us due to licensing restrictions).  For power estimation we used Mentor Graphics ModelSim to capture circuit activity and used that as input to Innovus. We use CACTI~\cite{cacti} to model the area and energy consumption of the on-chip shared SRAM memories which are divided into three chunks the AM, BM, and CM. 
We also use CACTI~\cite{cacti} to model the area and energy consumption of the SRAM scratchpads (SPs).
 Finally, we use Micron's DRAM model~\cite{micron} to estimate the energy consumption and latency of the off-chip memory.
\cref{tbl:arch_config} shows the default baseline and \OURL configurations. Both architectures compress zero values off-chip using the CompressingDMA method~\cite{rhu2018compressing}.

\begin{table}
\centering
\begin{tabular}{|l|l|l|l|}
\hline
\multicolumn{4}{|c|}{\textbf{\OURS and Baseline}} \\\hline 
Tile   & $4\times4$ PEs   & \# of Tiles        & 16 \\\hline
Total PEs &  256  & AM SRAM        &$256KB{\times}4$ Banks/Tile \\\hline
PE MACs/Cycle   & 16 FP32    & BM SRAM        & $256KB{\times}4$  Banks/Tile \\\hline
Total MACs/cycle  & 4096 & CM SRAM     & $256KB{\times}4$ Banks/Tile \\\hline
Staging Buff. Depth & 3 &   Scratchpads & $1KB{\times}3$ Banks each \\ \hline
Transposer Buff. & 1KB &   Transposers &  15 \\ \hline
Tech Node    & 65nm & Frequency    & 500 MHz \\\hline

     \multicolumn{2}{|c|}{Off-Chip Memory}
     & \multicolumn{2}{c|}{16GB 4-channel LPDDR4-3200} \\ \hline
\end{tabular}
\caption{Baseline and \OURL default configurations.}
\label{tbl:arch_config}
\end{table}

\subsection{Performance}\label{sec:performance_results}
\begin{figure}%
                \centering
                \includegraphics[width=0.48\textwidth]{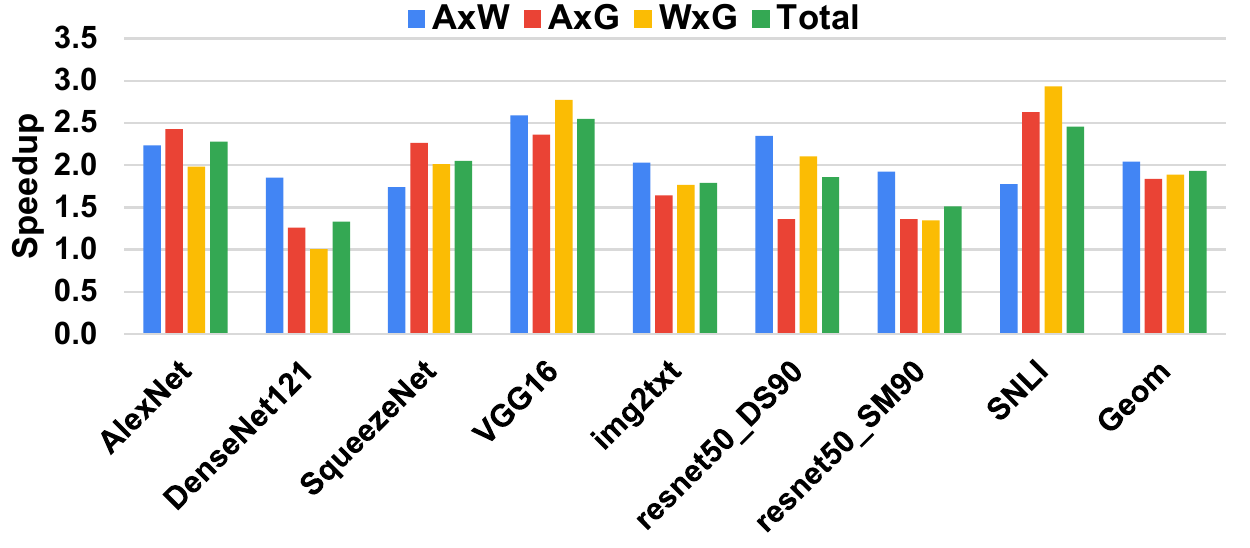}
                \caption{Speedup of \OURS over the baseline architecture.}
                \label{fig:speedup}
\end{figure}

\cref{fig:speedup} shows the speedup of \OURS over the baseline architecture for each model and for each of the three operations $A\star W$, $A\star G$ and $W\star G$. Since the amount of sparsity and its pattern in each of the tensors differs across models, layers and training phase, the speedup will be different per operation.  On average, \OURS achieves a speedup of $1.95{\times}$ over the baseline while it never slows down execution (for these measurements we do not power-gate any of the \OURL components ever). For DenseNet121 the speedup with \OURL for the third operation $W\star G$ is negligible. DenseNet121 uses a batch normalization layer between each convolution layer and the subsequent ReLU layer. This layer absorbs all the sparsity in the gradients. In addition, it is a dense model and thus has virtually no sparsity in the weights.

\subsection{Speedup Over Time}\label{sec:speedup_overTime}
\begin{figure}%
                \centering
                \includegraphics[width=0.48\textwidth]{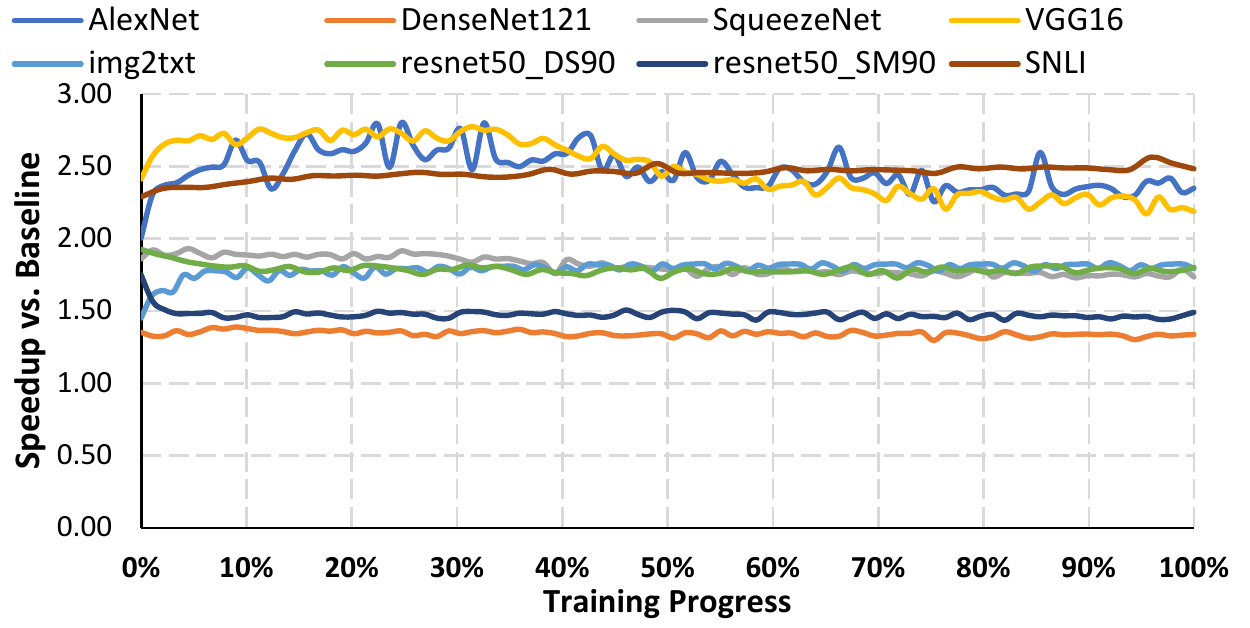}
                \caption{Speedup of \OURS as training progresses.}
                \label{fig:speedup_overTime}
\end{figure}
\cref{fig:speedup_overTime} shows the speedup of \OURS over the baseline as the training progresses from first epoch up until training converges. The speedups \OURS achieves are fairly stable throughout the entire training process. The measurements reveal two trends. For the ResNet50 models, which were trained with methods that induce model sparsity during training, the speedup is higher during the first few epochs and then it declines and stabilizes at around $5\%$ of the training epochs. For example, resnet50\_SM90 speedup starts at $1.75\times$ and then drops and settles at around $1.5\times$. Similar, albeit slightly more subdued behavior is seen for resnet50\_DS90 where speedup starts at $1.95\times$ and then stabilizes at $1.8\times$. This behavior is due to the pruning algorithm which starts by aggressively pruning many weights at the beginning which the training process then ``reclaims'' to recover the accuracy of the model. 

For the dense models, where most of the sparsity that \OURS exploits originates from the activations and gradients, the speedup tends to follow an overturned U-shape curve. This is especially pronounced for AlexNet and VGG16. The speedup starts low at the first epoch due to the random initialization of the model. Then speedup rapidly increases during the first few epochs as the model is quickly improving by learning what features of the input data are irrelevant for the task. This  translates to rapid increases in sparsity in the activations and the gradients. The speedup then stabilizes until $40\%-50\%$ of the training process is reached. It then gradually decreases as we enter the second half of the training process where the model starts to extract some of the less-important previously discarded features to improve accuracy. During the final quarter of the training process, the speedup stabilizes as the model parameters are very close to their optimal values and thus the sparsity of the activations and gradients is fairly stable. Rhu~\textit{et al.} have made similar observations when studying sparsity during training for the purpose of compressing data off-chip~\cite{rhu2018compressing}.

\subsection{Area Overhead, Power and Energy Efficiency}\label{sec:power}
\begin{figure}%
                \centering
                \includegraphics[width=0.5\textwidth]{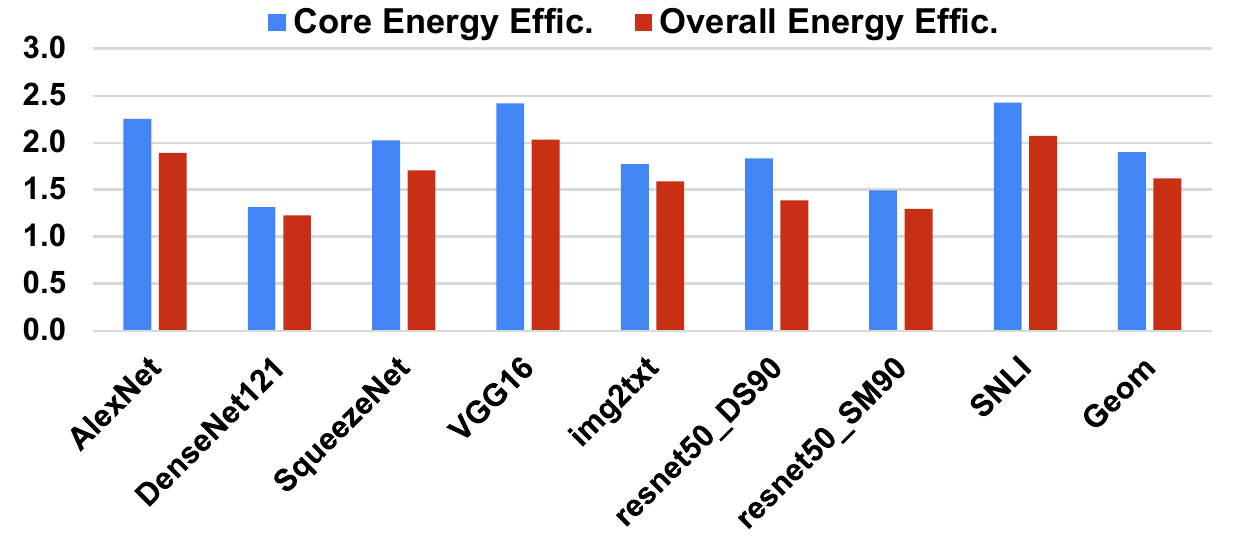}
                \caption{Energy efficiency of \OURS over the baseline.}
                \label{fig:energy_effic}
                \vspace{-10pt}
\end{figure}

\cref{tbl:area_power} shows a breakdown of the area and the power consumption for \OURS and the baseline. Even when the on-chip memory and off-chip DRAM are not taken into account, the area and power overheads of \OURS over the baseline are small. Only an $9\%$ extra silicon area and a $2\%$ power consumption overhead are needed for the schedulers and the back-end shufflers. However, given the speedup that \OURS achieves, the compute logic of \OURS is on average $1.89\times$ more energy efficient than the baseline. The per model and the overall average energy efficient measurements for the compute logic and the whole chip are reported in~\ref{fig:energy_effic}.

Each of the on-chip AM, BM, and CM memories would need $192\ mm^2$ of area whereas the scratchpads would need a total of $17\ mm^2$. In total when considering both compute and memory area for the whole chip, the area overhead of \OURS becomes imperceptible (1.0005$\times$). As~\cref{fig:energy_effic} shows, when we take the accesses to the on-chip memories, the scratchpads, and the off-chip DRAM into account, \OURS is still overall $1.6\times$ more energy efficient than the baseline.

\begin{figure}%
                \centering
                \includegraphics[width=0.48\textwidth]{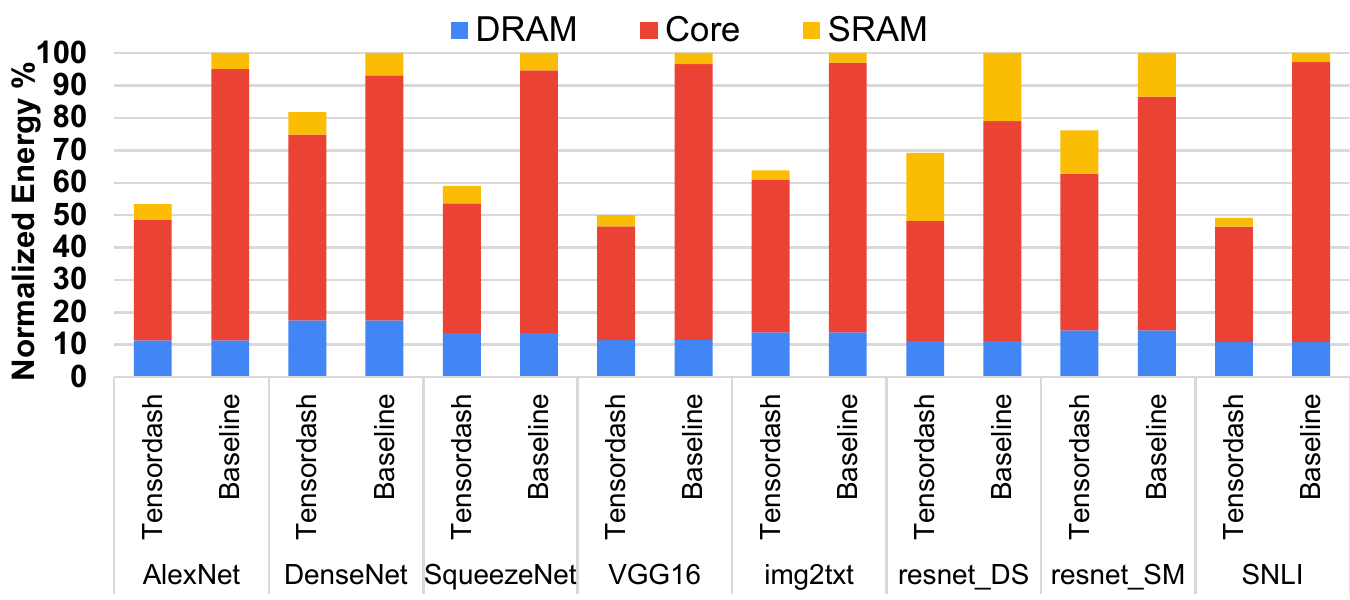}
                \caption{Energy consumption breakdown of \OURS and Baseline: off-chip DRAM, compute logic and on-chip SRAM.}
                \label{fig:energy_breakdown}
                \vspace{-10pt}
\end{figure}

\cref{fig:energy_breakdown} reports the energy consumed by \OURL relative to the baseline. The measurements also show a breakdown of the energy consumed across three main components: the off-chip data transfers, core logic, and the on-chip memory modules. \OURS significantly reduces the energy consumption of the core which dominates the energy consumption of the system.

 \begin{table}
                \centering\scriptsize

                \caption{Area [$mm^2$] and Power consumption [$mW$] breakdown of \OURS vs. Baseline. On-chip AM/BM/CM and scratchpad are not included.}
                \label{tbl:area_power}
                \begin{tabularx}{\linewidth}{|X|c|c||c|c|} \hline
                    &  \multicolumn{2}{c||}{\textbf{Area ($mm^2$)}} & \multicolumn{2}{c|}{\textbf{Power ($mW$)}} \\ \hhline{~----}
                \multirow{2}{*} {} & \textbf{\OURS} & \textbf{Baseline} & \textbf{\OURS} & \textbf{Baseline} \\ \hline \hline
                         
            \textbf{Compute Cores} & \multicolumn{2}{c||}{30.41} & \multicolumn{2}{c|}{13,910}   \\ \hline
            \textbf{Transposers} & \multicolumn{2}{c||}{0.38} & \multicolumn{2}{c|}{47.3} \\ \hline
            \textbf{Schedulers+B-Side MUXes} & 0.91 & - & 102.8 & -   \\ \hline
            \textbf{A-Side MUXes} & 1.73 & - & 145.3 & -\\ \hline
            \textbf{Total} & \textbf{33.44}& \textbf{30.80} & 14,205 & 13,957\\ \hline
            \textbf{Normalized} & $\mathbf{1.09\times}$ & $\mathbf{1\times}$ & $\mathbf{1.02\times}$ & $\mathbf{1\times}$ \\ \hline
            \multicolumn{1}{|c}{}&\multicolumn{2}{r|}{\textbf{Energy  Efficiency}}  & $\mathbf{1.89\times}$ & $\mathbf{1\times}$ \\ \hline
            
                \end{tabularx}
        \end{table}

\subsection{Analysis}\label{sec:scalability_rows}

\noindent\textbf{$\bullet$ Tile Geometry: } We study the performance behavior of the \OURL PE when it is used to compose tiles. For this purpose we vary the number of PE rows and columns per tile and study how this affects performance. As the tile geometry changes stalls will occur due to inter-PE synchronization which in turn is caused by work imbalance.

\begin{figure}%
                \centering
                \includegraphics[width=0.4\textwidth]{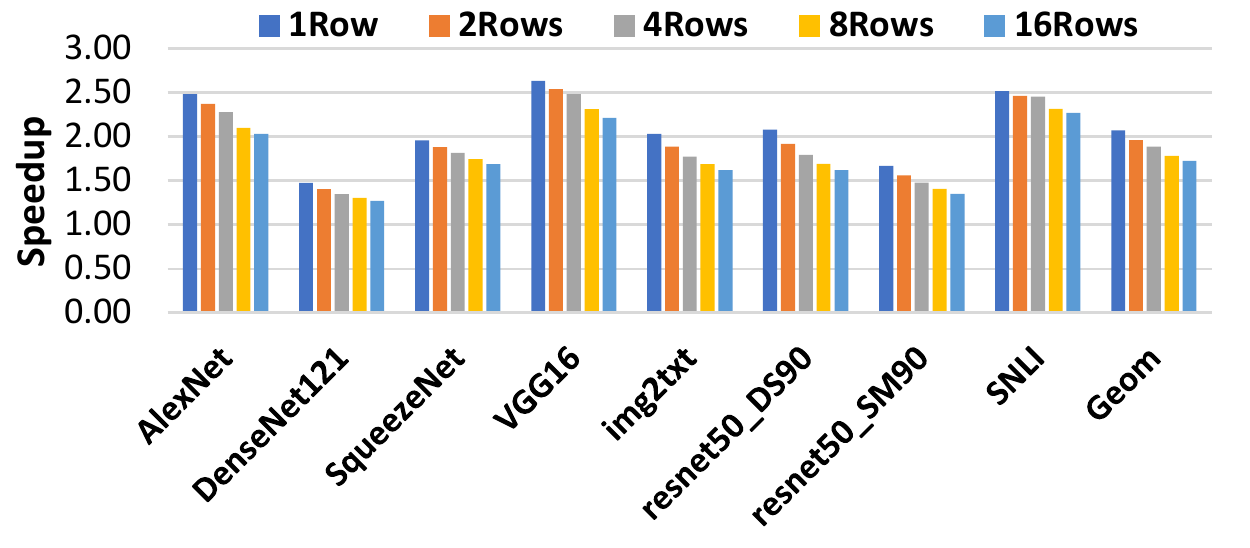}
                \caption{\OURS speedup vs. number of PE rows.}
                \label{fig:scalability_rows}
                                \vspace{-10pt}

\end{figure}

\noindent\textbf{\textit{Rows:} }\cref{fig:scalability_rows} shows how performance varies with various configurations of \OURL where the number of rows is varied from 1 and up to 16 (the number of columns is fixed at 4). The average speedup decreases from $2.1\times$ for a tile with 1 row to  $1.72\times$ when the tile has 16 rows. Since all PEs have to wait for the slowest one, the more rows the more frequent stalls due to work imbalance will occur. As we scale up the number of rows per tile, the data values that are concurrently processed exhibit density imbalance across rows. This can stall some rows since \textit{all} have to wait for the one with the densest value stream. In effect, as the number of rows increases, it becomes less likely that scheduling such a large group of values will result in skipping the entire processing cycle and advancing to the next group. The main reason why this occurs is that the non-zero activations and gradients tend to cluster in certain 2D feature maps whereas the other 2D maps become more sparse. This clustering phenomenon is fundamental in such models especially towards the deeper layers where each filter is trained to extract specific high level features. In other words, an input sample having a feature $X$ and lacking a feature $Y$ would typically exhibit a dense map corresponding to the former and a sparse for the latter.  This phenomenon is more pronounced for $A{\times}G$, the second backward convolution, where the 2D feature maps of the activations and the gradients are convolved.

\begin{figure}%
                \centering
                \includegraphics[width=0.40\textwidth]{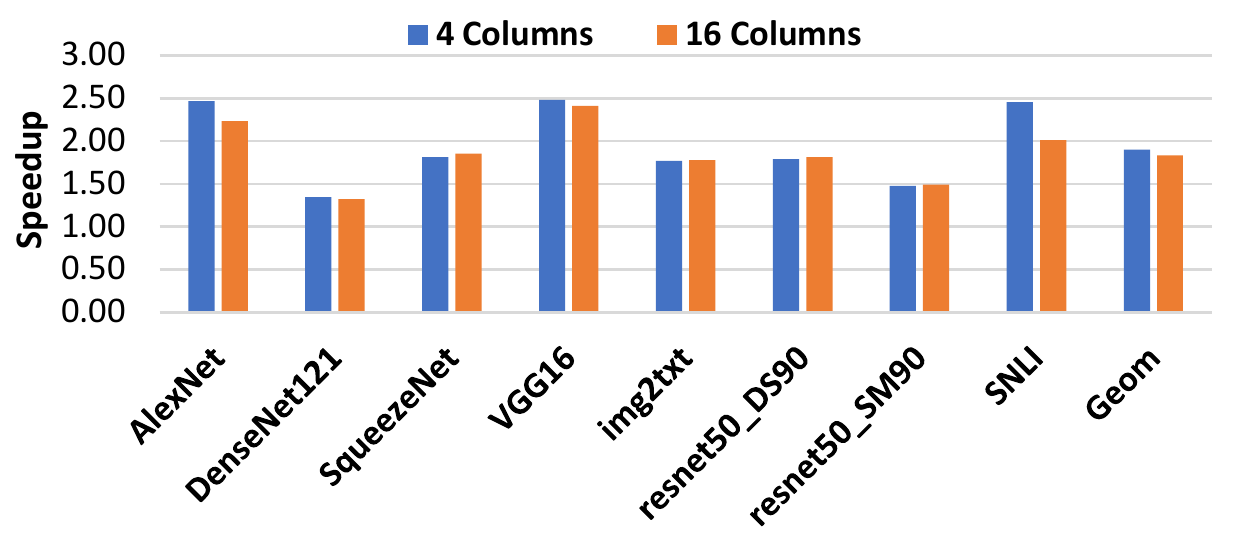}
                \caption{\OURL speedup vs. PE columns.}
                \label{fig:scalability_columns}
                                                \vspace{-10pt}

\end{figure}

\noindent\textbf{\textit{Columns: }}Figure~\ref{fig:scalability_columns} shows how the speedup achieved by \OURL scales as we vary instead the number of columns per tile from 4 to 16 (the number of rows stays at 4). This effectively scales the maximum throughput to 16K MACs per cycle. Since in this configuration studied we exploit sparsity only on one side, increasing the number of columns does not affect performance as much. All rows still have to wait for the row with the most work. However, increasing the columns allows us to process more windows in parallel while sharing the same schedule across the rows. Slight drops in performance are due predominantly to fragmentation due to layer dimensions.

\begin{figure}%
                \centering
                \includegraphics[width=0.48\textwidth]{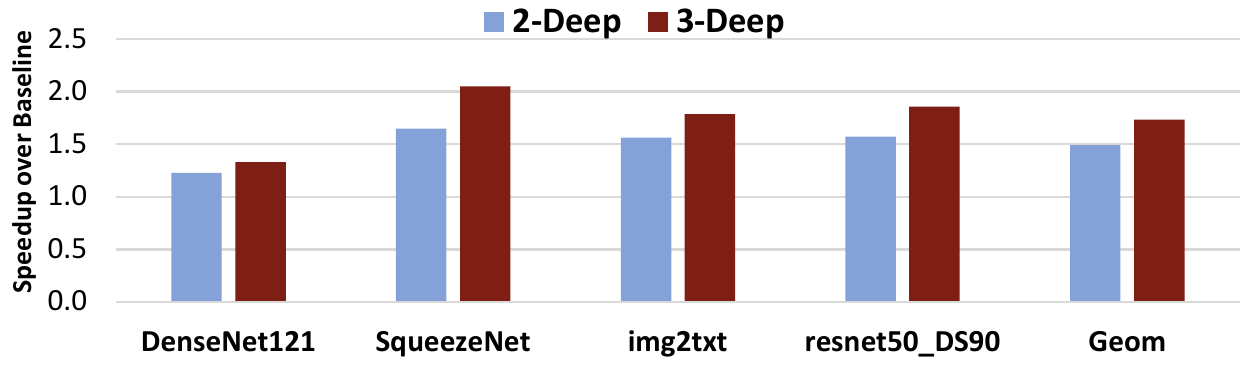}
                \caption{\OURL speedup for staging depth of 2 vs 3.}
                \label{fig:staging_depth}
                                \vspace{-10pt}

\end{figure}
\begin{figure}%
                \centering
                \includegraphics[width=0.48\textwidth]{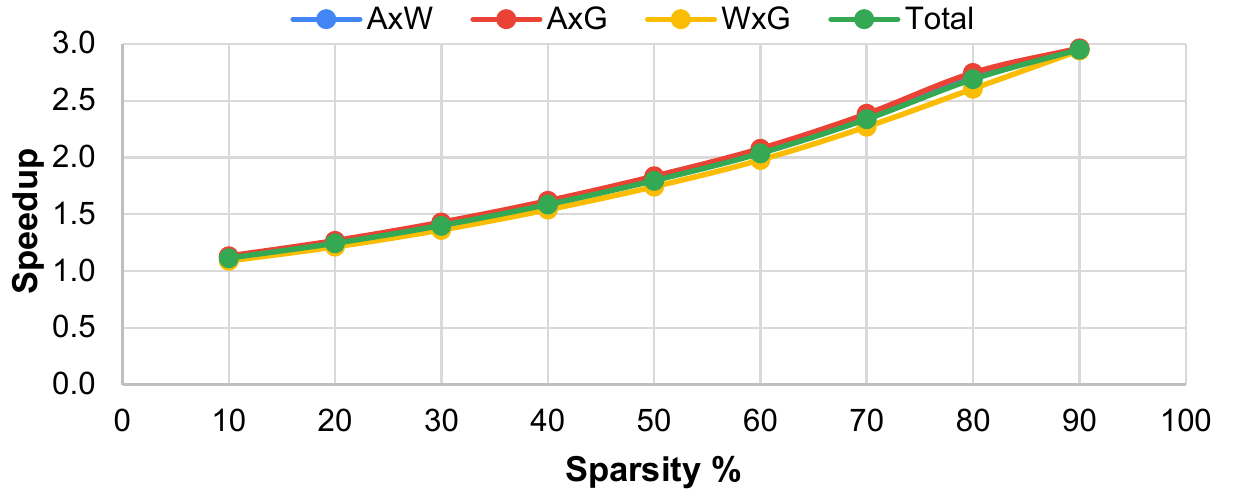}
                \caption{\OURL speedup for randomly sparse tensors.}
                \label{fig:random_sparsity}
                \vspace{-10pt}
\end{figure}

\noindent\textbf{$\bullet$ Staging Buffer Depth/Lookahead: }Figure~\ref{fig:staging_depth} reports speedups with \OURL with 2-deep staging buffers (lookahead of 1); 5 movements per multiplier. This is a lower-cost configuration. While speedups are lower, they are still considerable representing another appealing cost vs. performance design point.

\noindent\textbf{$\bullet$ Effect of Tensor Sparsity: }
To determine whether \OURL remains effective regardless of the sparsity structure of the input tensors, we experimented with synthetically generated sparse tensors with sparsity levels ranging from $10\%$ up to $90\%$.  We used the architecture of the third conv. layer from DenseNet121 but populated the tensors  using randomly generated values. For each level of sparsity (0.1 to 0.9 with step 0.1) we generated 10 samples of inputs. We then performed all three operations for each sample using these randomly generated tensors. We report the average across all samples for a given sparsity level (the deviation across samples was below 5\%).
As \cref{fig:random_sparsity} shows, performance with \OURS closely follows the amount of sparsity in the input. Recall that given the 3-deep staging buffers we use, the maximum possible speedup with \OURL even if the tensor contains only zeros is $3\times$. The figure shows that when the ideal speedup is below $3\times$ \OURL comes close to what is ideally possible. For example, with $10\%$ sparsity, an optimal machine would be $1.11\times$ faster assuming all the ineffectual MACs are eliminated. \OURS is approximately $1.1\times$ speedup. For $90\%$ sparsity, an ideal machine would be able to achieve a $10\times$ speedup. However, due to the limited depth of the staging buffer, \OURL would ideally be $3\times$ faster. The experiment shows that \OURS comes close to what is ideally possible. It is $2.95\times$ faster. The speedups are consistent across the forward and backward operations.

\noindent\textbf{$\bullet$ Training with Bfloat16: }\label{sec:scalability_bf16}
Recent research work showed that deep neural networks could be trained using narrower floating-point data types such as bfloat16~\cite{bfloat16,bfloat16_2}. Mixed-precision training using standard FP16 and FP32 has also been  shown to be successful~\cite{nvidia_mixedP}. We implemented \OURS and baseline configurations that use bfloat16 arithmetic. Even when we consider only the compute logic, our synthesis+layout results show that the area and power consumption overheads of \OURS vs. the baseline are $1.13\times$ and $1.05\times$. The overheads are higher but still low. The various components scale differently as the data type shrinks: Some, such as the priority encoders, do not scale. Others, such as the zero comparators, scale linearly. Finally, the multiplier cores scale nearly quadratically. However, when the scaled-down on-chip memory structures are taken into account, the area overhead is nearly the same as it was for the FP32 configuration and stands at $1.0005\times$. In terms of energy efficiency, the compute logic of \OURS would still be on average $1.84\times$ more energy efficient than the baseline. When accesses to the on-chip and the off-chip memory are taken into account, \OURS is overall $1.43\times$ more energy efficient.

\noindent\textbf{$\bullet$ A Model with Virtually No Sparsity: } We experimented with GCN~\cite{GCNNfb}, a natural language processing model which we trained on the Wikitext-2 dataset~\cite{wikidataset}. It exhibits virtually no sparsity. Still, \OURL improves performance by 1\% since a few layers exhibit about 5\% sparsity. Without power-gating \OURL overall energy efficiency is 0.5\% lower than the baseline.
\section{Related Work}\label{sec:related}
The architecture of choice for training has been the graphics processor which a good fit for data-parallel computations. 
Neural networks and GPUs have evolved almost symbiotically during the last few years with GPUs introducing features to aid inference and training~\cite{DBLP:conf/hotchips/Burgess19}. XeonPhi is another architecture that is well suited to this type of data-parallel workload~\cite{Xeonphi}. However, there have been designs that target explicitly machine learning training. \textit{Here we review just a few. We regret that due to space limitations it is not possible to refer to them all (note to reviewers: we do plan to revise for the final version given an extra page, e.g., Habana, Graphcore, Cerebras, etc.).  }

Scaledeep is a scalable architecture for training. It utilizes heterogenous tiles and chips, an optimized network topology, low-overhead hardware-assisted synchronization, and optimized model partitioning~\cite{ScaleDeep}. DaDianNao is one of the earliest accelerator architectures targeting primarily inference, whose tiles however, could be fused to support 32b arithmetic for training~\cite{DaDiannao}. Newer version of the TPU also support training~\cite{TPU}. Plasticine does not target machine learning exclusively but a wide set of parallel computation patterns which include those needed for stochastic gradient descent~\cite{placts}. Caterpillar provide hierarchical support for collective communication semantics to  provides the flexibility needed to efficiently training various networks with both stochastic and batched gradient descent based techniques~\cite{Li_2017}. NXT is a near-memory accelerator comprising several general purpose cores and specialized co-processors targeting both inference and training~\cite{nearmemarch}. Intel's NNP-T  (Spring Crest) supports both FP32 and FP16~\cite{DBLP:conf/hotchips/Yang19}.  It uses a stack of 4 8GB HMB2-2400 external memories, 60MB of on-chip memory.

\OURL proposes a processing element that can exploit sparsity and which can be used to compose tiles. As such it is not meant as a competitor for the overall accelerator architecture. That said, in every case there will be several considerations that need close attention and evaluation.



\section{Conclusion}
\label{sec:conclusion}


As we discussed in the introduction, training is an exascale problem at the datacenter. It is also one that will need to be supported for certain applications at the edge. This work is valuable for such efforts as it presented a low-level processing element that could be of value for building accelerators for either segment. While there is a multitude of options and configurations that are worthwhile exploring their interaction with \OURL, we believe that this work is sufficient and stands on its own. It does demonstrate a practical use and serves as motivation for such studies.

Given the importance of training there is a large and ever increasing volume of works for accelerating training in software, hardware or both. We commented on a subset of these methods in the introduction. While \OURL will interact with several of these training acceleration methods, it is at first-order complementary with many since it operates at the very low level of the MAC units. Which is to say that we believe that our method can be of value as a replacement PE for several existing hardware accelerators and in conjunction with several existing software-level training accelerations techniques. Demonstrating this requires further work. Nevertheless, this work has made the necessary step of establishing that such investigations are worthwhile. Specifically, this work has established clearly that our method can indeed deliver benefits and thus serves to motivate such investigations.

\bibliographystyle{IEEEtran.bst}
\bibliography{ref}

\begin{thebibliography}{10}
\providecommand{\url}[1]{#1}
\csname url@samestyle\endcsname
\providecommand{\newblock}{\relax}
\providecommand{\bibinfo}[2]{#2}
\providecommand{\BIBentrySTDinterwordspacing}{\spaceskip=0pt\relax}
\providecommand{\BIBentryALTinterwordstretchfactor}{4}
\providecommand{\BIBentryALTinterwordspacing}{\spaceskip=\fontdimen2\font plus
\BIBentryALTinterwordstretchfactor\fontdimen3\font minus
  \fontdimen4\font\relax}
\providecommand{\BIBforeignlanguage}[2]{{%
\expandafter\ifx\csname l@#1\endcsname\relax
\typeout{** WARNING: IEEEtran.bst: No hyphenation pattern has been}%
\typeout{** loaded for the language `#1'. Using the pattern for}%
\typeout{** the default language instead.}%
\else
\language=\csname l@#1\endcsname
\fi
#2}}
\providecommand{\BIBdecl}{\relax}
\BIBdecl

\bibitem{ScaleDeep}
\BIBentryALTinterwordspacing
S.~Venkataramani, A.~Ranjan, S.~Banerjee, D.~Das, S.~Avancha, A.~Jagannathan,
  A.~Durg, D.~Nagaraj, B.~Kaul, P.~Dubey, and A.~Raghunathan, ``Scaledeep: A
  scalable compute architecture for learning and evaluating deep networks,'' in
  \emph{Proceedings of the 44th Annual International Symposium on Computer
  Architecture}, ser. ISCA '17.\hskip 1em plus 0.5em minus 0.4em\relax New
  York, NY, USA: ACM, 2017. [Online]. Available:
  \url{http://doi.acm.org/10.1145/3079856.3080244}
\BIBentrySTDinterwordspacing

\bibitem{TPU}
\BIBentryALTinterwordspacing
N.~P. Jouppi, C.~Young, N.~Patil, D.~Patterson, G.~Agrawal, R.~Bajwa, S.~Bates,
  S.~Bhatia, N.~Boden, A.~Borchers, R.~Boyle, P.-l. Cantin, C.~Chao, C.~Clark,
  J.~Coriell, M.~Daley, M.~Dau, J.~Dean, B.~Gelb, T.~V. Ghaemmaghami,
  R.~Gottipati, W.~Gulland, R.~Hagmann, C.~R. Ho, D.~Hogberg, J.~Hu, R.~Hundt,
  D.~Hurt, J.~Ibarz, A.~Jaffey, A.~Jaworski, A.~Kaplan, H.~Khaitan,
  D.~Killebrew, A.~Koch, N.~Kumar, S.~Lacy, J.~Laudon, J.~Law, D.~Le, C.~Leary,
  Z.~Liu, K.~Lucke, A.~Lundin, G.~MacKean, A.~Maggiore, M.~Mahony, K.~Miller,
  R.~Nagarajan, R.~Narayanaswami, R.~Ni, K.~Nix, T.~Norrie, M.~Omernick,
  N.~Penukonda, A.~Phelps, J.~Ross, M.~Ross, A.~Salek, E.~Samadiani, C.~Severn,
  G.~Sizikov, M.~Snelham, J.~Souter, D.~Steinberg, A.~Swing, M.~Tan,
  G.~Thorson, B.~Tian, H.~Toma, E.~Tuttle, V.~Vasudevan, R.~Walter, W.~Wang,
  E.~Wilcox, and D.~H. Yoon, ``In-datacenter performance analysis of a tensor
  processing unit,'' in \emph{Proceedings of the 44th Annual International
  Symposium on Computer Architecture}, ser. ISCA '17.\hskip 1em plus 0.5em
  minus 0.4em\relax New York, NY, USA: ACM, 2017. [Online]. Available:
  \url{http://doi.acm.org/10.1145/3079856.3080246}
\BIBentrySTDinterwordspacing

\bibitem{ML_co2}
\BIBentryALTinterwordspacing
E.~Strubell, A.~Ganesh, and A.~McCallum, ``Energy and policy considerations for
  deep learning in {NLP},'' \emph{CoRR}, vol. abs/1906.02243, 2019. [Online].
  Available: \url{http://arxiv.org/abs/1906.02243}
\BIBentrySTDinterwordspacing

\bibitem{dist_training1}
\BIBentryALTinterwordspacing
J.~Dean, G.~S. Corrado, R.~Monga, K.~Chen, M.~Devin, Q.~V. Le, M.~Z. Mao,
  M.~Ranzato, A.~Senior, P.~Tucker, K.~Yang, and A.~Y. Ng, ``Large scale
  distributed deep networks,'' in \emph{Proceedings of the 25th International
  Conference on Neural Information Processing Systems - Volume 1}, ser.
  NIPS'12.\hskip 1em plus 0.5em minus 0.4em\relax USA: Curran Associates Inc.,
  2012. [Online]. Available:
  \url{http://dl.acm.org/citation.cfm?id=2999134.2999271}
\BIBentrySTDinterwordspacing

\bibitem{dist_training2}
\BIBentryALTinterwordspacing
R.~Mayer and H.~Jacobsen, ``Scalable deep learning on distributed
  infrastructures: Challenges, techniques and tools,'' \emph{CoRR}, vol.
  abs/1903.11314, 2019. [Online]. Available:
  \url{http://arxiv.org/abs/1903.11314}
\BIBentrySTDinterwordspacing

\bibitem{dist_training3}
J.~Dean, G.~Corrado, R.~Monga, K.~Chen, M.~Devin, M.~Mao, M.~Ranzato,
  A.~Senior, P.~Tucker, K.~Yang \emph{et~al.}, ``Large scale distributed deep
  networks,'' in \emph{Advances in neural information processing systems},
  2012.

\bibitem{eyeriss}
Y.-H. Chen, J.~Emer, and V.~Sze, ``Eyeriss: A spatial architecture for
  energy-efficient dataflow for convolutional neural networks,'' in \emph{ACM
  SIGARCH Computer Architecture News}, vol.~44, no.~3.\hskip 1em plus 0.5em
  minus 0.4em\relax IEEE Press, 2016.

\bibitem{TernGrad}
\BIBentryALTinterwordspacing
W.~Wen, C.~Xu, F.~Yan, C.~Wu, Y.~Wang, Y.~Chen, and H.~Li, ``Terngrad: Ternary
  gradients to reduce communication in distributed deep learning,'' in
  \emph{Advances in Neural Information Processing Systems 30}, I.~Guyon, U.~V.
  Luxburg, S.~Bengio, H.~Wallach, R.~Fergus, S.~Vishwanathan, and R.~Garnett,
  Eds.\hskip 1em plus 0.5em minus 0.4em\relax Curran Associates, Inc., 2017,
  pp. 1509--1519. [Online]. Available:
  \url{http://papers.nips.cc/paper/6749-terngrad-ternary-gradients-to-reduce-communication-in-distributed-deep-learning.pdf}
\BIBentrySTDinterwordspacing

\bibitem{UCNN}
\BIBentryALTinterwordspacing
K.~Hegde, J.~Yu, R.~Agrawal, M.~Yan, M.~Pellauer, and C.~W. Fletcher, ``Ucnn:
  Exploiting computational reuse in deep neural networks via weight
  repetition,'' in \emph{Proceedings of the 45th Annual International Symposium
  on Computer Architecture}, ser. ISCA '18.\hskip 1em plus 0.5em minus
  0.4em\relax Piscataway, NJ, USA: IEEE Press, 2018. [Online]. Available:
  \url{https://doi.org/10.1109/ISCA.2018.00062}
\BIBentrySTDinterwordspacing

\bibitem{GIST}
\BIBentryALTinterwordspacing
A.~Jain, A.~Phanishayee, J.~Mars, L.~Tang, and G.~Pekhimenko, ``Gist: Efficient
  data encoding for deep neural network training,'' in \emph{Proceedings of the
  45th Annual International Symposium on Computer Architecture}, ser. ISCA
  '18.\hskip 1em plus 0.5em minus 0.4em\relax Piscataway, NJ, USA: IEEE Press,
  2018. [Online]. Available: \url{https://doi.org/10.1109/ISCA.2018.00070}
\BIBentrySTDinterwordspacing

\bibitem{bfloat16Google}
\BIBentryALTinterwordspacing
S.~Wang and P.~Kanwar, ``Bfloat16: The secret to high performance on cloud
  tpus,'' 2019. [Online]. Available:
  \url{https://cloud.google.com/blog/products/ai-machine-learning/bfloat16-the-secret-to-high-performance-on-cloud-tpus}
\BIBentrySTDinterwordspacing

\bibitem{bfloat16}
\BIBentryALTinterwordspacing
D.~D. Kalamkar, D.~Mudigere, N.~Mellempudi, D.~Das, K.~Banerjee, S.~Avancha,
  D.~T. Vooturi, N.~Jammalamadaka, J.~Huang, H.~Yuen, J.~Yang, J.~Park,
  A.~Heinecke, E.~Georganas, S.~Srinivasan, A.~Kundu, M.~Smelyanskiy, B.~Kaul,
  and P.~Dubey, ``A study of {BFLOAT16} for deep learning training,''
  \emph{CoRR}, vol. abs/1905.12322, 2019. [Online]. Available:
  \url{http://arxiv.org/abs/1905.12322}
\BIBentrySTDinterwordspacing

\bibitem{bfloat16_2}
Google, ``Using bfloat16 with tensorflow models,''
  \url{https://cloud.google.com/tpu/docs/bfloat16}.

\bibitem{DBLP:conf/iclr/0002MMKAB0VKGHD18}
\BIBentryALTinterwordspacing
D.~Das, N.~Mellempudi, D.~Mudigere, D.~D. Kalamkar, S.~Avancha, K.~Banerjee,
  S.~Sridharan, K.~Vaidyanathan, B.~Kaul, E.~Georganas, A.~Heinecke, P.~Dubey,
  J.~Corbal, N.~Shustrov, R.~Dubtsov, E.~Fomenko, and V.~O. Pirogov, ``Mixed
  precision training of convolutional neural networks using integer
  operations,'' in \emph{6th International Conference on Learning
  Representations, {ICLR} 2018, Vancouver, BC, Canada, April 30 - May 3, 2018,
  Conference Track Proceedings}, 2018. [Online]. Available:
  \url{https://openreview.net/forum?id=H135uzZ0-}
\BIBentrySTDinterwordspacing

\bibitem{Koster:2017:FAN:3294771.3294937}
\BIBentryALTinterwordspacing
U.~K\"{o}ster, T.~J. Webb, X.~Wang, M.~Nassar, A.~K. Bansal, W.~H. Constable,
  O.~H. Elibol, S.~Gray, S.~Hall, L.~Hornof, A.~Khosrowshahi, C.~Kloss, R.~J.
  Pai, and N.~Rao, ``Flexpoint: An adaptive numerical format for efficient
  training of deep neural networks,'' in \emph{Proceedings of the 31st
  International Conference on Neural Information Processing Systems}, ser.
  NIPS'17.\hskip 1em plus 0.5em minus 0.4em\relax USA: Curran Associates Inc.,
  2017. [Online]. Available:
  \url{http://dl.acm.org/citation.cfm?id=3294771.3294937}
\BIBentrySTDinterwordspacing

\bibitem{mixedP}
\BIBentryALTinterwordspacing
P.~Micikevicius, S.~Narang, J.~Alben, G.~F. Diamos, E.~Elsen, D.~Garc{\'{\i}}a,
  B.~Ginsburg, M.~Houston, O.~Kuchaiev, G.~Venkatesh, and H.~Wu, ``Mixed
  precision training,'' in \emph{6th International Conference on Learning
  Representations, {ICLR} 2018, Vancouver, BC, Canada, April 30 - May 3, 2018,
  Conference Track Proceedings}, 2018. [Online]. Available:
  \url{https://openreview.net/forum?id=r1gs9JgRZ}
\BIBentrySTDinterwordspacing

\bibitem{nvidia_mixedP}
NVIDIA, ``Training with mixed precision,''
  \url{https://docs.nvidia.com/deeplearning/sdk/mixed-precision-training/index.html}.

\bibitem{Drumond:2018:TDH:3326943.3326985}
\BIBentryALTinterwordspacing
M.~Drumond, T.~Lin, M.~Jaggi, and B.~Falsafi, ``Training dnns with hybrid block
  floating point,'' in \emph{Proceedings of the 32Nd International Conference
  on Neural Information Processing Systems}, ser. NIPS'18.\hskip 1em plus 0.5em
  minus 0.4em\relax USA: Curran Associates Inc., 2018. [Online]. Available:
  \url{http://dl.acm.org/citation.cfm?id=3326943.3326985}
\BIBentrySTDinterwordspacing

\bibitem{HALP}
C.~De~Sa, M.~Leszczynski, J.~Zhang, A.~Marzoev, C.~R. Aberger, K.~Olukotun, and
  C.~R{\'e}, ``High-accuracy low-precision training,'' \emph{arXiv preprint
  arXiv:1803.03383}, 2018.

\bibitem{dynamic_sparse_reparam}
H.~Mostafa and X.~Wang, ``Parameter efficient training of deep convolutional
  neural networks by dynamic sparse reparameterization,'' in
  \emph{International Conference on Machine Learning}, 2019.

\bibitem{eager_pruning}
\BIBentryALTinterwordspacing
J.~Zhang, X.~Chen, M.~Song, and T.~Li, ``Eager pruning: Algorithm and
  architecture support for fast training of deep neural networks,'' in
  \emph{Proceedings of the 46th International Symposium on Computer
  Architecture}, ser. ISCA '19.\hskip 1em plus 0.5em minus 0.4em\relax New
  York, NY, USA: ACM, 2019. [Online]. Available:
  \url{http://doi.acm.org/10.1145/3307650.3322263}
\BIBentrySTDinterwordspacing

\bibitem{DropBack}
M.~Golub, G.~Lemieux, and M.~Lis, ``Dropback: Continuous pruning during
  training,'' \emph{arXiv preprint arXiv:1806.06949}, 2018.

\bibitem{PACT}
J.~Choi, Z.~Wang, S.~Venkataramani, P.~I.-J. Chuang, V.~Srinivasan, and
  K.~Gopalakrishnan, ``Pact: Parameterized clipping activation for quantized
  neural networks,'' \emph{arXiv preprint arXiv:1805.06085}, 2018.

\bibitem{LQ-nets}
D.~Zhang, J.~Yang, D.~Ye, and G.~Hua, ``Lq-nets: Learned quantization for
  highly accurate and compact deep neural networks,'' in \emph{Proceedings of
  the European Conference on Computer Vision (ECCV)}, 2018.

\bibitem{MeProp}
\BIBentryALTinterwordspacing
X.~Sun, X.~Ren, S.~Ma, and H.~Wang, ``meprop: Sparsified back propagation for
  accelerated deep learning with reduced overfitting,'' in \emph{Proceedings of
  the 34th International Conference on Machine Learning - Volume 70}, ser.
  ICML'17.\hskip 1em plus 0.5em minus 0.4em\relax JMLR.org, 2017. [Online].
  Available: \url{http://dl.acm.org/citation.cfm?id=3305890.3306022}
\BIBentrySTDinterwordspacing

\bibitem{rhu2018compressing}
M.~Rhu, M.~O'Connor, N.~Chatterjee, J.~Pool, Y.~Kwon, and S.~W. Keckler,
  ``Compressing dma engine: Leveraging activation sparsity for training deep
  neural networks,'' in \emph{2018 IEEE International Symposium on High
  Performance Computer Architecture (HPCA)}.\hskip 1em plus 0.5em minus
  0.4em\relax IEEE, 2018.

\bibitem{narrowFP_ibm}
\BIBentryALTinterwordspacing
N.~Wang, J.~Choi, D.~Brand, C.-Y. Chen, and K.~Gopalakrishnan, ``Training deep
  neural networks with 8-bit floating point numbers,'' in \emph{Proceedings of
  the 32Nd International Conference on Neural Information Processing Systems},
  ser. NIPS'18.\hskip 1em plus 0.5em minus 0.4em\relax USA: Curran Associates
  Inc., 2018. [Online]. Available:
  \url{http://dl.acm.org/citation.cfm?id=3327757.3327866}
\BIBentrySTDinterwordspacing

\bibitem{eyeriss-jssc}
Y.-H. Chen, T.~Krishna, J.~Emer, and V.~Sze, ``Eyeriss: An energy-efficient
  reconfigurable accelerator for deep convolutional neural networks,''
  \emph{IEEE Journal of Solid-State Circuits}, vol.~52, Jan 2017.

\bibitem{cambricon:2016}
S.~Liu, Z.~Du, J.~Tao, D.~Han, T.~Luo, Y.~Xie, Y.~Chen, and T.~Chen,
  ``Cambricon: An instruction set architecture for neural networks,'' in
  \emph{2016 {IEEE/ACM} {Intl'} {Conf.} on {Computer} {Architecture} ({ISCA})},
  2016.

\bibitem{CambriconX}
\BIBentryALTinterwordspacing
S.~Zhang, Z.~Du, L.~Zhang, H.~Lan, S.~Liu, L.~Li, Q.~Guo, T.~Chen, and Y.~Chen,
  ``Cambricon-x: An accelerator for sparse neural networks,'' in \emph{Intl'
  Symp. on Microarchitecture}, 2016. [Online]. Available:
  \url{https://doi.org/10.1109/MICRO.2016.7783723}
\BIBentrySTDinterwordspacing

\bibitem{SCNN}
A.~Parashar, M.~Rhu, A.~Mukkara, A.~Puglielli, R.~Venkatesan, B.~Khailany,
  J.~Emer, S.~W. Keckler, and W.~J. Dally, ``{SCNN:} an accelerator for
  compressed-sparse convolutional neural networks,'' in \emph{Intl' Symp. on
  Computer Architecture}, ser. ISCA '17, 2017.

\bibitem{CambriconS:MICRO2018}
X.~Zhou, Z.~Du, Q.~Guo, C.~Liu, C.~Wang, X.~Zhou, L.~Li, T.~Chen, and Y.~Chen,
  ``{Cambricon-S:} addressing irregularity in sparse neural networks through a
  cooperative software/hardware approach,'' in \emph{Intl' Symp. on
  Microarchitecture}, 2018.

\bibitem{Tactical}
\BIBentryALTinterwordspacing
A.~Delmas~Lascorz, P.~Judd, D.~M. Stuart, Z.~Poulos, M.~Mahmoud, S.~Sharify,
  M.~Nikolic, K.~Siu, and A.~Moshovos, ``Bit-tactical: A software/hardware
  approach to exploiting value and bit sparsity in neural networks,'' in
  \emph{Proceedings of the Twenty-Fourth International Conference on
  Architectural Support for Programming Languages and Operating Systems}, ser.
  ASPLOS '19.\hskip 1em plus 0.5em minus 0.4em\relax New York, NY, USA: ACM,
  2019. [Online]. Available: \url{http://doi.acm.org/10.1145/3297858.3304041}
\BIBentrySTDinterwordspacing

\bibitem{CambriconXMICRO16}
S.~Zhang, Z.~Du, L.~Zhang, H.~Lan, S.~Liu, L.~Li, Q.~Guo, T.~Chen, and Y.~Chen,
  ``Cambricon-x: An accelerator for sparse neural networks,'' in \emph{Intl'
  Symp. on Microarchitecture}, 2016.

\bibitem{Cnvlutin2}
\BIBentryALTinterwordspacing
P.~Judd, A.~D. Lascorz, S.~Sharify, and A.~Moshovos, ``Cnvlutin2:
  Ineffectual-activation-and-weight-free deep neural network computing,''
  \emph{CoRR}, vol. abs/1705.00125, 2017. [Online]. Available:
  \url{http://arxiv.org/abs/1705.00125}
\BIBentrySTDinterwordspacing

\bibitem{SparTen}
\BIBentryALTinterwordspacing
A.~Gondimalla, N.~Chesnut, M.~Thottethodi, and T.~N. Vijaykumar, ``Sparten: A
  sparse tensor accelerator for convolutional neural networks,'' in
  \emph{Proceedings of the 52Nd Annual IEEE/ACM International Symposium on
  Microarchitecture}, ser. MICRO '52.\hskip 1em plus 0.5em minus 0.4em\relax
  New York, NY, USA: ACM, 2019. [Online]. Available:
  \url{http://doi.acm.org/10.1145/3352460.3358291}
\BIBentrySTDinterwordspacing

\bibitem{han_eie:isca_2016}
S.~Han, X.~Liu, H.~Mao, J.~Pu, A.~Pedram, M.~A. Horowitz, and W.~J. Dally,
  ``Eie: Efficient inference engine on compressed deep neural network,'' in
  \emph{Intl' Symp. on Computer Architecture}, 2016.

\bibitem{imagenet}
O.~Russakovsky, J.~Deng, H.~Su, J.~Krause, S.~Satheesh, S.~Ma, Z.~Huang,
  A.~Karpathy, A.~Khosla, M.~Bernstein, A.~C. Berg, and L.~Fei-Fei,
  ``{ImageNet} {Large} {Scale} {Visual} {Recognition} {Challenge},''
  \emph{CoRR}, vol. abs/1409.0575, Sep. 2014.

\bibitem{alexnet}
\BIBentryALTinterwordspacing
A.~Krizhevsky, I.~Sutskever, and G.~E. Hinton, ``Imagenet classification with
  deep convolutional neural networks,'' \emph{Commun. ACM}, vol.~60, May 2017.
  [Online]. Available: \url{http://doi.acm.org/10.1145/3065386}
\BIBentrySTDinterwordspacing

\bibitem{densenet121}
\BIBentryALTinterwordspacing
G.~Huang, Z.~Liu, and K.~Q. Weinberger, ``Densely connected convolutional
  networks,'' \emph{CoRR}, vol. abs/1608.06993, 2016. [Online]. Available:
  \url{http://arxiv.org/abs/1608.06993}
\BIBentrySTDinterwordspacing

\bibitem{SqueezeNet}
\BIBentryALTinterwordspacing
F.~N. Iandola, M.~W. Moskewicz, K.~Ashraf, S.~Han, W.~J. Dally, and K.~Keutzer,
  ``Squeezenet: Alexnet-level accuracy with 50x fewer parameters and
  {\textless}1mb model size,'' \emph{CoRR}, vol. abs/1602.07360, 2016.
  [Online]. Available: \url{http://arxiv.org/abs/1602.07360}
\BIBentrySTDinterwordspacing

\bibitem{vgg}
K.~Simonyan and A.~Zisserman, ``Very deep convolutional networks for
  large-scale image recognition,'' \emph{arXiv preprint arXiv:1409.1556}, 2014.

\bibitem{resnet}
\BIBentryALTinterwordspacing
K.~He, X.~Zhang, S.~Ren, and J.~Sun, ``Deep residual learning for image
  recognition,'' \emph{CoRR}, vol. abs/1512.03385, 2015. [Online]. Available:
  \url{http://arxiv.org/abs/1512.03385}
\BIBentrySTDinterwordspacing

\bibitem{img2txt}
\BIBentryALTinterwordspacing
O.~Vinyals, A.~Toshev, S.~Bengio, and D.~Erhan, ``Show and tell: Lessons
  learned from the 2015 {MSCOCO} image captioning challenge,'' \emph{CoRR},
  vol. abs/1609.06647, 2016. [Online]. Available:
  \url{http://arxiv.org/abs/1609.06647}
\BIBentrySTDinterwordspacing

\bibitem{snli}
\BIBentryALTinterwordspacing
S.~R. Bowman, G.~Angeli, C.~Potts, and C.~D. Manning, ``A large annotated
  corpus for learning natural language inference,'' in \emph{Proceedings of the
  2015 Conference on Empirical Methods in Natural Language Processing}.\hskip
  1em plus 0.5em minus 0.4em\relax Lisbon, Portugal: Association for
  Computational Linguistics, Sep. 2015. [Online]. Available:
  \url{https://www.aclweb.org/anthology/D15-1075}
\BIBentrySTDinterwordspacing

\bibitem{dynSparse}
H.~Mostafa and X.~Wang, ``Parameter efficient training of deep convolutional
  neural networks by dynamic sparse reparameterization,'' in
  \emph{International Conference on Machine Learning}, 2019.

\bibitem{sparseMom}
T.~Dettmers and L.~Zettlemoyer, ``Sparse networks from scratch: Faster training
  without losing performance,'' \emph{arXiv preprint arXiv:1907.04840}, 2019.

\bibitem{synopsys_site}
{Synopsys}, ``{Design Compiler},''
  {http://www.synopsys.com/Tools/\\Implementation/RTLSynthesis/DesignCompiler/Pages},
  2019.

\bibitem{innovus}
Cadence, ``Innovus implementation system,''
  \url{https://www.cadence.com/content/cadence-www/global/en_US/home/tools/digital-design-and-signoff/hierarchical-design-and-floorplanning/innovus-implementation-system.html}.

\bibitem{cacti}
HewlettPackard, ``Cacti,'' \url{https://github.com/HewlettPackard/cacti}.

\bibitem{micron}
I.~Micron~Technology, ``Ddr4 power calculator 4.0,''
  \url{https://www.micron.com/~/media/documents/products/power-calculator/ddr4_power_calc.xlsm}.

\bibitem{GCNNfb}
\BIBentryALTinterwordspacing
Y.~N. Dauphin, A.~Fan, M.~Auli, and D.~Grangier, ``Language modeling with gated
  convolutional networks,'' in \emph{Proceedings of the 34th International
  Conference on Machine Learning - Volume 70}, ser. ICML'17.\hskip 1em plus
  0.5em minus 0.4em\relax JMLR.org, 2017. [Online]. Available:
  \url{http://dl.acm.org/citation.cfm?id=3305381.3305478}
\BIBentrySTDinterwordspacing

\bibitem{wikidataset}
\BIBentryALTinterwordspacing
S.~Merity, C.~Xiong, J.~Bradbury, and R.~Socher, ``Pointer sentinel mixture
  models,'' in \emph{5th International Conference on Learning Representations,
  {ICLR} 2017, Toulon, France, April 24-26, 2017, Conference Track
  Proceedings}, 2017. [Online]. Available:
  \url{https://openreview.net/forum?id=Byj72udxe}
\BIBentrySTDinterwordspacing

\bibitem{DBLP:conf/hotchips/Burgess19}
\BIBentryALTinterwordspacing
J.~Burgess, ``{RTX} {ON} - the {NVIDIA} {TURING} {GPU},'' in \emph{2019 {IEEE}
  Hot Chips 31 Symposium (HCS), Cupertino, CA, USA, August 18-20, 2019}, 2019.
  [Online]. Available: \url{https://doi.org/10.1109/HOTCHIPS.2019.8875651}
\BIBentrySTDinterwordspacing

\bibitem{Xeonphi}
R.~Rahman, \emph{Intel Xeon Phi Coprocessor Architecture and Tools: The Guide
  for Application Developers}, 1st~ed.\hskip 1em plus 0.5em minus 0.4em\relax
  Berkely, CA, USA: Apress, 2013.

\bibitem{DaDiannao}
Y.~Chen, T.~Luo, S.~Liu, S.~Zhang, L.~He, J.~Wang, L.~Li, T.~Chen, Z.~Xu,
  N.~Sun, and O.~Temam, ``{DaDianNao: A Machine-Learning Supercomputer},'' in
  \emph{Intl' Symp. on Microarchitecture}, 2014.

\bibitem{placts}
R.~{Prabhakar}, Y.~{Zhang}, D.~{Koeplinger}, M.~{Feldman}, T.~{Zhao},
  S.~{Hadjis}, A.~{Pedram}, C.~{Kozyrakis}, and K.~{Olukotun}, ``Plasticine: A
  reconfigurable architecture for parallel patterns,'' in \emph{2017 ACM/IEEE
  44th Annual International Symposium on Computer Architecture (ISCA)}, June
  2017.

\bibitem{Li_2017}
\BIBentryALTinterwordspacing
Y.~Li and A.~Pedram, ``Caterpillar: Coarse grain reconfigurable architecture
  for accelerating the training of deep neural networks,'' \emph{2017 IEEE 28th
  International Conference on Application-specific Systems, Architectures and
  Processors (ASAP)}, Jul 2017. [Online]. Available:
  \url{http://dx.doi.org/10.1109/ASAP.2017.7995252}
\BIBentrySTDinterwordspacing

\bibitem{nearmemarch}
F.~{Schuiki}, M.~{Schaffner}, F.~K. {Gürkaynak}, and L.~{Benini}, ``A scalable
  near-memory architecture for training deep neural networks on large in-memory
  datasets,'' \emph{IEEE Transactions on Computers}, vol.~68, April 2019.

\bibitem{DBLP:conf/hotchips/Yang19}
\BIBentryALTinterwordspacing
A.~Yang, ``Deep learning training at scale spring crest deep learning
  accelerator (intel{\textregistered} nervana{\texttrademark} {NNP-T)},'' in
  \emph{2019 {IEEE} Hot Chips 31 Symposium (HCS), Cupertino, CA, USA, August
  18-20, 2019}, 2019. [Online]. Available:
  \url{https://doi.org/10.1109/HOTCHIPS.2019.8875643}
\BIBentrySTDinterwordspacing

\end{thebibliography}

\pagebreak


\end{document}